\documentclass[fleqn,usenatbib]{mnras}

\usepackage{newtxtext,newtxmath}

\usepackage[T1]{fontenc}

\DeclareRobustCommand{\VAN}[3]{#2}
\let\VANthebibliography\thebibliography
\def\thebibliography{\DeclareRobustCommand{\VAN}[3]{##3}\VANthebibliography}

\usepackage{graphicx}	
\usepackage{amsmath}	
\graphicspath{{./}{figures/}}
\usepackage{threeparttable}
\newcommand{\insight}{\textit{Insight}-HXMT}
\newcommand{\fermi}{\textit{Fermi}/GBM}

\title[Burst Search with Poisson likelihood ratio]{Burst search method based on likelihood ratio in Poisson Statistics}

\author[C. Cai et al.]{
Ce Cai$^{1,2,3}$,
Shao-Lin Xiong$^{2}$\thanks{E-mail: xiongsl@ihep.ac.cn},
Wang-Chen Xue$^{2,3}$,
Yi Zhao$^{2,4}$,
Shuo Xiao$^{2,3,9,10}$,
Qi-Bin Yi$^{2,5}$,
\newauthor
Zhi-Wei Guo$^{2,6}$,
Jia-Cong Liu$^{2,3}$,
Yan-Qiu Zhang$^{2,3}$,
Chao Zheng$^{2,3}$,
Sheng-Lun Xie$^{2,7}$,
Yan-Qi Du$^{2,8}$,
\newauthor
Xiao-Yun Zhao$^{2}$,
Cheng-Kui Li$^{2}$,
Ping Wang$^{2}$,
Wen-Xi Peng$^{2}$,
Shi-Jie Zheng$^{2,3}$,
Li-Ming Song$^{2,3}$,
\newauthor
Xin-Qiao Li$^{2}$,
Xiang-Yang Wen$^{2}$,
Fan Zhang$^{2}$
\\
$^{1}$ College of Physics, Hebei Normal University, 20 South Erhuan Road, Shijiazhuang 050024, Hebei, China\\ 
$^{2}$ Key Laboratory of Particle Astrophysics, Institute of High Energy
Physics, Chinese Academy of Sciences, Beijing 100049, China\\
$^{3}$ University of Chinese Academy of Sciences, Chinese Academy of Sciences, Beijing 100049, China\\
$^{4}$ Department of Astronomy, Beijing Normal University, Beijing 100875, China\\
$^{5}$ School of Physics and Optoelectronics, Xiangtan University, Xiangtan 411105, Hunan, China \\
$^{6}$ College of physics Sciences Technology, Hebei University, No. 180 Wusi Dong Road, Lian Chi District, Baoding 071002, Hebei, China\\
$^{7}$ Institute of Astrophysics, Central China Normal University, Wuhan 430079, Hubei, China \\
$^{8}$ Information Science and Technology, Southwest Jiaotong University, Chengdu 610031, Sichuan, China\\
$^{9}$ Guizhou Provincial Key Laboratory of Radio Astronomy and Data Processing, Guizhou Normal University, Guiyang 550001, GuiZhou, China\\
$^{10}$ School of Physics and Electronic Science, Guizhou Normal University, Guiyang 550001, GuiZhou, China\\}

\date{Accepted XXX. Received YYY; in original form ZZZ}

\pubyear{2022}

\begin{document}
\label{firstpage}
\pagerange{\pageref{firstpage}--\pageref{lastpage}}
\maketitle
\begin{abstract}
Searching for X-ray and gamma-ray bursts, including Gamma-ray bursts (GRBs), Soft Gamma-ray Repeaters (SGRs) and high energy transients associated with Gravitational wave (GW) events or Fast radio bursts (FRBs), is of great importance in the multi-messenger and multi-wavelength era. Although a coherent search based on the likelihood ratio and Gaussian statistics has been established and utilized in many studies, this Gaussian-based method could be problematic for weak and short bursts which usually have very few counts.
To deal with all bursts including weak ones, here we propose the coherent search in Poisson statistics. We studied the difference between Poisson-based and Gaussian-based search methods by Monte Carlo simulations, and find
that the Poisson-based search method has advantages compared to the Gaussian case especially for weak bursts. 
Our results show that, for very weak bursts with very low number of counts, the Poisson-based search can provide higher significance than the Gaussian-based search,
and its likelihood ratio (for background fluctuation) still generally follows the $\chi^{2}$ distribution, making the significance estimation of searched bursts very convenient. Thus, we suggest that the coherent search based on Poisson likelihood ratio is more appropriate in the search for generic transients including very weak ones.

\end{abstract}




\section{Introduction}  \label{introducion}
Recent discoveries of a Gamma-ray burst (GRB 170817A) associated with the Gravitational wave (GW 170817) \citep{TheLIGOScientific:2017qsa,Goldstein:2017mmi, Savchenko:2017ffs,Li:2017iup} and a non-thermal X-ray burst from the Galactic magnetar (SGR J1935+2154) associated with the Fast radio burst (FRB 200428) \citep{2021NatAs...5..378L,2020Natur.587...54C,bochenek2020fast,Mereghetti_2020,tavani2020xray,2021NatAs...5..372R} highlight the importance of the observations of high energy transients in X-ray and gamma-ray band. 

As the ground-based gravitational wave observatories (LIGO, Virgo and KAGRA) continue to upgrade, their detected GW events would likely be further and the presumptive gamma-ray bursts associated with those GW events would probably be very weak. 
Also, as most of the detected FRBs are of extragalactic origin \citep[e.g.][]{2007Sci...318..777L},
if they are also associated with X-ray bursts from extragalactic magnetars \citep[e.g.][]{2020ApJ...899..106Y}, then these X-ray bursts are expected to be very weak and short. 
In fact, gamma-ray transients with short duration have been conceived to be the possible counterparts of FRBs \citep{Guidorzi_2020}. For example, \cite{Guidorzi_2020,2020Guidorzi} searched \insight\,data for FRB-associated gamma-ray counterparts with time scales down to milliseconds and even sub-milliseconds (i.e. 0.1 ms). 

On the other hand, the bright and sharp peak of GRBs may appear in very short duration down to milliseconds due to the tip-of-the-iceberg effect \citep{2016ApJS..227....7L,2022ApJ...927..157M}.
It is also well known that GRBs tend to be shorter in duration in higher energy range band \citep[e.g.][]{2022ApJS..259...46S}. 
It is found that the duration of magnetar short bursts could be shorter than about 50 ms \citep{cai2022insighthxmt,2015ApJS..218...11C,Lin_2020b}. As \cite{2005Natur.434.1098H} pointed out, an extragalactic giant flar as bright as SGR1806–20 could appear as very short-duration depending on its distance owing to the tip-of-the-iceberg effect.
In fact, much shorter (down to millisecond) bursts have been detected from the Earth \citep[i.e. Terrestrial Gamma-ray flashes and Terrestrial Electron Beams,][]{1994STIN...9611316F,articleXiong}.

We note that most gamma-ray detectors usually have temporal resolution of several microseconds, which is sufficient to detect sub-milliseconds bursts. For example, GECAM \citep{LiXinQiao2022,AnZhengHua2022} has a temporal resolution of about 0.1 $\mu$s \citep{10.1093/mnras/stac085} which allow us to exploit some extreme short bursts. These short and weak GRB and SGR bursts would leave a small number of signal counts in detectors.

Meanwhile, during such a short duration of the burst, the background counts would be very small as well.
For example, the average background level of each GRD detector of GECAM and each NaI detector of \fermi\,is about 1 count per 1 ms (in energy range of about 10-200 keV) \citep{2011AIPC.1358..313L,2022MNRAS.tmp..994X}. For each CsI detector of \insight/HE, this number is about 0.6 counts per 1 ms (in energy range of about 80 - 800 keV) \citep{Zhang_2020,2020SCPMA..6349503L}. 

Therefore, it is critically important to find and study these weak X-ray and gamma-ray bursts in the multi-messenger and multi-wavelength era. However, caution should be made in the analysis of weak bursts with very few counts.
In fact, there are many dedicated studies dealing with weak signal and low counts statistics \citep{1983ApJ...272..317L,2017A&A...605A..51K,1999NIMPA.431..239H,2001NIMPA.457..384H}.

The first step of studying weak bursts is finding them by burst search algorithm. In an earlier work, \cite{Blackburn:2014rqa} developed a sensitive coherent search method for targeted \fermi\, \citep{Meegan_2009} follow-up observation of GW events. This method has been used to follow up all LIGO triggers including sub-threshold GW triggers. It is also used to search for GRBs and magnetar bursts \citep{2021GCN.30125....1F,2021GCN.30140....1C}. 
Several improvements have been made to this coherent search method by many authors \citep{Goldstein:2016zfh,Goldstein:2019tfz,10.1093/mnras/stab2760}, including background estimation, spectral template, calculation speed, search sensitivity as well as rejection of false triggers caused by instrumental effects. 
However, this coherent search method is deduced with Gaussian statistics, which is a good approximation for bright bursts with sufficient number of counts, but may be problematic for weak bursts with very few counts.



Motivated by the generic search for all kinds of bursts, including the weak bursts, 
here we deduced the coherent search method based on maximum likelihood ratio in Poisson statistics. Then we compared the performance of the Poisson-based search method to the Gaussian-based one by detailed Monte Carlo simulations. 
In section \ref{sec:method}, we describe the likelihood ratio based coherent search method in both Gaussian and Poisson statistics. Section \ref{sec:simulation and result} depicts the simulation and result. Finally, discussions and conclusions are presented in section \ref{sec:Discussions} and \ref{sec:CONCLUSIONS}.

\begin{figure*}
    \includegraphics[width=1\textwidth]{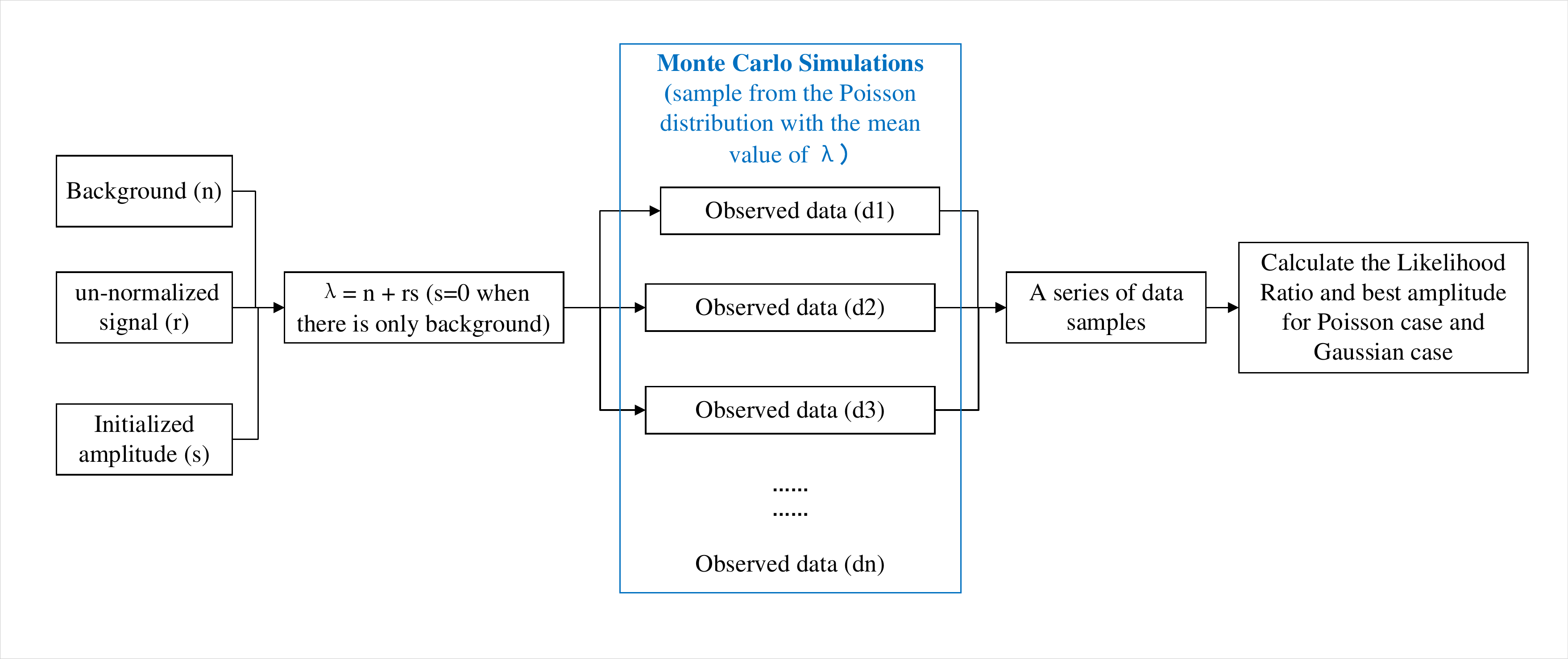}
    \caption{Illustration of the framework of our simulation.}
    \label{fig:simulate_flowchart}
\end{figure*}

\section{Burst Search Methods} \label{sec:method}
Burst search methods discussed here are based on the framework of likelihood ratio, which was first presented by Neyman and Pearson \citep{10.1093/biomet/20A.3-4.263}. The likelihood ratio is defined as:

\begin{equation}
\Lambda = \frac{ P(d|H_{1}(s))}{ P(d|H_{0})}, \label{LR_def}
\end{equation}

where $P(d|H_{1}(s))$ and $P(d|H_{0})$ are the likelihood functions for two hypothesis and models, $H_{1}(s)$ and $H_{0}$, respectively. $d$ and $s$ represent observation data and model parameters, respectively. For simplicity, one usually use log-likelihood ratio ($\mathcal L$) which is defined as: $\mathcal L= {\rm ln} \,\Lambda$. 

\cite{Blackburn:2014rqa} developed a coherent search method based on the likelihood ratio to search GBM data for various bursts \citep{2021GCN.30125....1F,2021GCN.30140....1C}. The likelihood ratio is defined as the ratio between probabilities (likelihood) of two hypothesis: (1) $H_{1}$: the observed counts are contributed by the background plus burst signal; (2) $H_{0}$: the observed data is purely background without any burst. This method has been improved by a series of studies \citep{Goldstein:2016zfh,Goldstein:2019tfz,Kocevski:2018suj} and widely used in the burst search of \fermi,\,\insight\,and other instruments \citep[e.g.][]{2020ApJ...893..100H,10.1093/mnras/stab2760}. 

In above studies, the likelihood function is formulated in the Gaussian statistics with the assumption of large number of counts. To deal with weak bursts with few counts, the likelihood can be rewritten in the Poisson statistics. Both the Gaussian and Poisson cases are presented as following.

\subsection{Gaussian case} \label{Gaussian data}
\label{sec:Gaussian} 
Following the presentation in \cite{Blackburn:2014rqa} and \cite{10.1093/mnras/stab2760}, the coherent search method based on the Gaussian statistics could be formulated as:

\begin{equation}
{{P}({d}_{i}|{H}_{1})}= \prod_{i=1}^j \frac{1}{\sqrt{2\pi}{\sigma}_{{d}_{i}}}\rm exp(-\frac{({\widetilde{d}_{i}-{r}_{i}s)^2}}{2\sigma^{2}_{{d}_{i}}}) \,\,(i=1,2...j), \label{eq1}
\end{equation}

\begin{equation}
{{P}({d}_{i}|{H}_{0})}= \prod_{i=1}^j \frac{1}{\sqrt{2\pi}{\sigma}_{{n}_{i}}}\rm exp(-\frac{{\widetilde{d}_{i}}^2}{2\sigma^{2}_{{n}_{i}}}), \label{eq2}
\end{equation}

\begin{equation}
\widetilde{d}_{i}={d}_{i}-\hat{n}_{i},
\end{equation}
where $i$ represents the number of data sets in each detectors and channels, $j$ is the total number of detectors and channels,
${d}_{i}$ and ${\sigma}_{{d}_{i}}$ are the observed data (counts) and standard deviation of the expected data (background+signal), respectively, $\hat{n}_{i}$ and ${\sigma}_{{n}_{i}}$ are the estimated background and the standard deviation of the background data, respectively, ${r}_{i}$ and $s$ represent the burst signal with default amplitude of 1 (i.e. un-normalized signal) and the intrinsic source amplitude, respectively,  $\widetilde{d}_{i}$ is the background-subtracted data (i.e. net counts).

Then we can define the log-likelihood ratio (hereafter likelihood ratio for simplicity):
\begin{equation}
{ \mathcal L_{\rm g} } = {\rm ln} \frac{{P}({d}_{i}|{H}_{1})}{{P}({d}_{i}|{H}_{0})} = \sum_{i=1}^j[{\rm ln}\frac{{\sigma}_{{n}_{i}}}{{\sigma}_{{d}_{i}}} + \frac{{\widetilde{d}_{i}}^2}{2\sigma^{2}_{{n}_{i}}} - \frac{({\widetilde{d}_{i}-{r}_{i}s)^2}}{2\sigma^{2}_{{d}_{i}}}].   \label{eq3}
\end{equation}
where $\mathcal L_{\rm g}$ is the likelihood ratio for Gaussian statistics.

\subsection{Poisson case}
\label{sec:Poisson} 

For the case of weak bursts with fewer counts, the calculation of likelihood should be based on the Poisson distribution, thus Eq.\ref{eq1}, \ref{eq2} and \ref{eq3} could be rewritten as: 

\begin{equation}
{{P}({d}_{i}|{H}_{1})}= \prod_{i=1}^j\frac{(r_{i}s+\hat{n}_{i})^{d_{i}}}{d_{i}!}\rm exp(-(r_{i}s+\hat{n}_{i}))
\end{equation}

\begin{equation}
{{P}({d}_{i}|{H}_{0})}= \prod_{i=1}^j\frac{(\hat{n}_{i})^{d_{i}}}{d_{i}!}\rm exp(-\hat{n}_{i})
\end{equation}

\begin{equation}
{ \mathcal L_{\rm p} } = {\rm ln} \frac{{P}({d}_{i}|{H}_{1})}{{P}({d}_{i}|{H}_{0})} = \sum_{i=1}^j[d_{i}\rm ln(\frac{r_{i}s+\hat{n}_{i}}{\hat{n}_{i}})-r_{i}s], \label{eq4}
\end{equation}
where $\mathcal L_{\rm p}$ represent the likelihood ratio of Poisson distribution. Other parameters are defined the same as section \ref{Gaussian data}.

As we know, when the mean values ($\mu$) is large, the Poisson distribution can be well approximated by a Gaussian distribution with standard deviation of $\sqrt{\mu}$. 
Also, the central limit theorem states that a variable should follow Gaussian distribution if it is the sum of many random variables that are independent identically distributed. Therefore, these two methods based on Poisson and Gaussian distribution may show significant difference only for those cases that the mean value is small and the number of variables that contribute to the sum of likelihood ratio is not large. 

\begin{figure*}
    \begin{tabular}{cc}
	\includegraphics[width=\columnwidth]{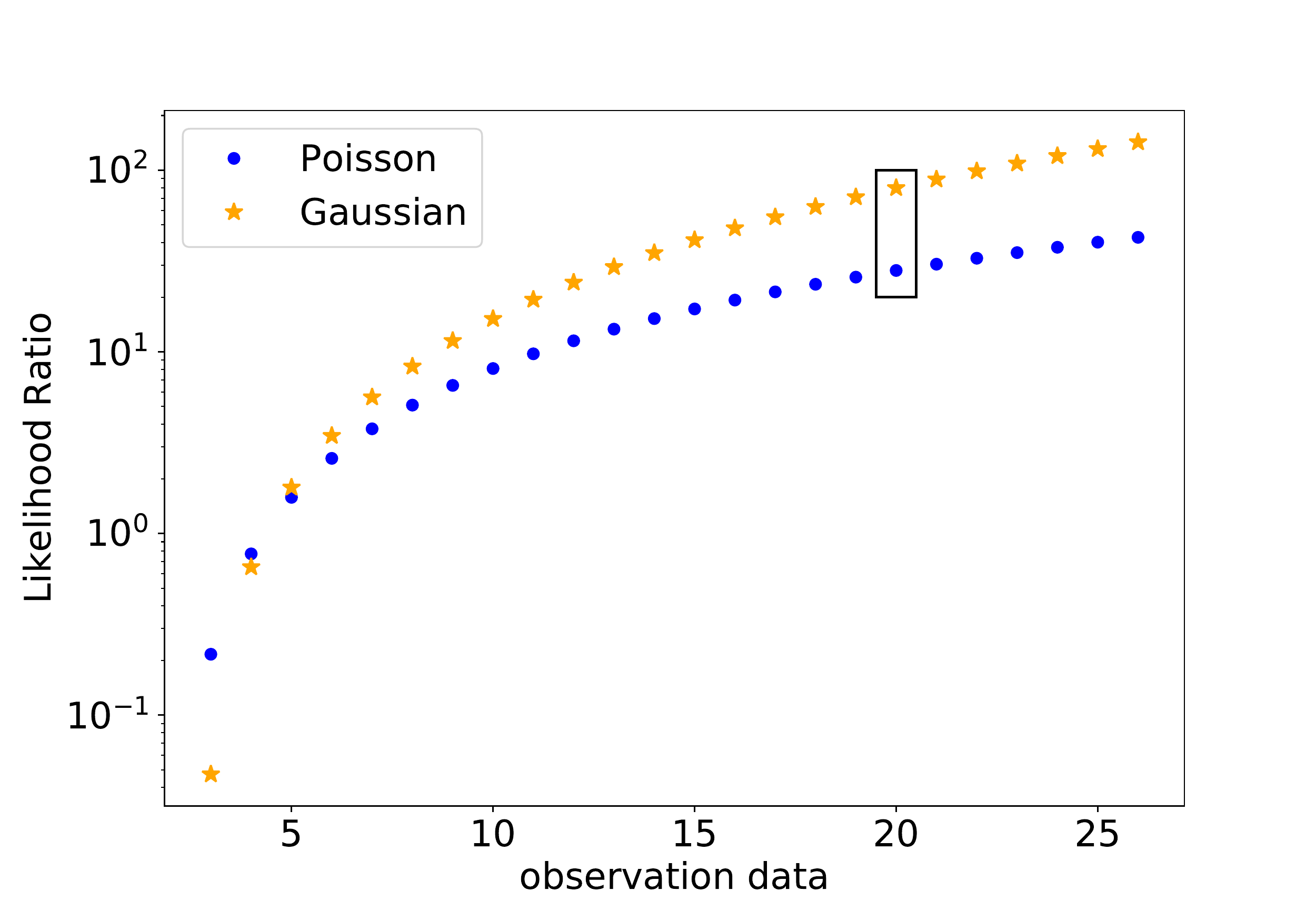} &
	\includegraphics[width=\columnwidth]{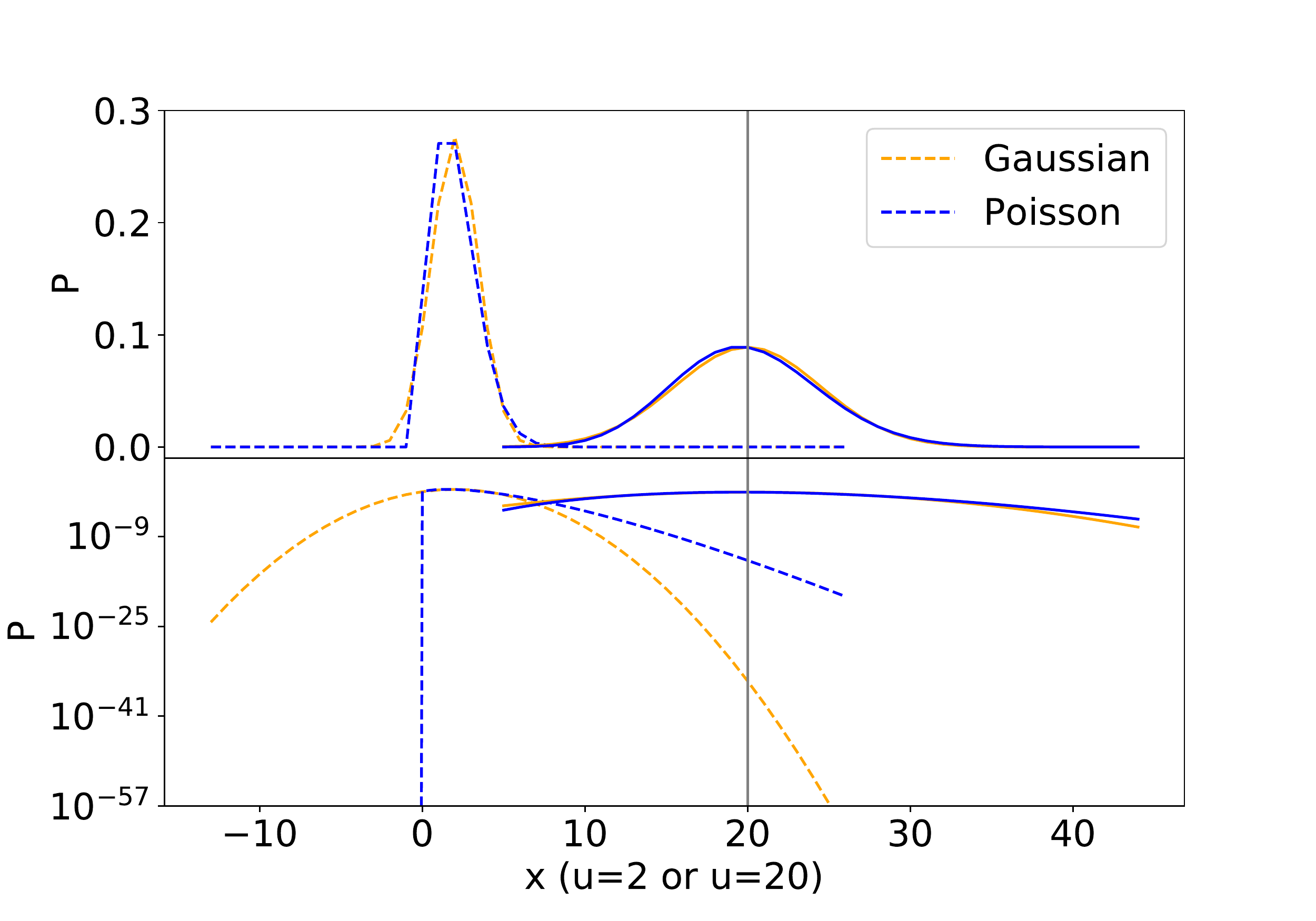} \\
	\end{tabular}
    \caption{
    \textit{Left}: Likelihood ratio based on Poisson statistics (blue points) and Gaussian statistics (orange stars) versus observed data counts. The expected value of background is set to 2 counts. The observed data is set to the mean value of background plus the mean value of signal, which is used to avoid the error of statistical fluctuations. The black box is the observed data of 20 counts. \textit{Right}: Poisson distribution (blue) versus Gaussian distribution (orange). The dashed lines and solid lines represent the expected counts of 2 and 20, respectively. The gray line is the observed data of 20 counts. The probabilities for Gaussian case is similar to that for Poisson case at high observed counts (e.g., 20 counts) and high expected counts (e.g., 20 counts), as shown in the top panel. The probabilities 
    for Gaussian case is different from that for Poisson case at high observed counts (e.g., 20 counts) and low expected counts (e.g., 2 counts), as shown in the lower panel (dashed orange line and blue line in the logarithm scale).}
    \label{fig:LR_counts}
\end{figure*}

\begin{figure*}
    \includegraphics[width=1\textwidth]{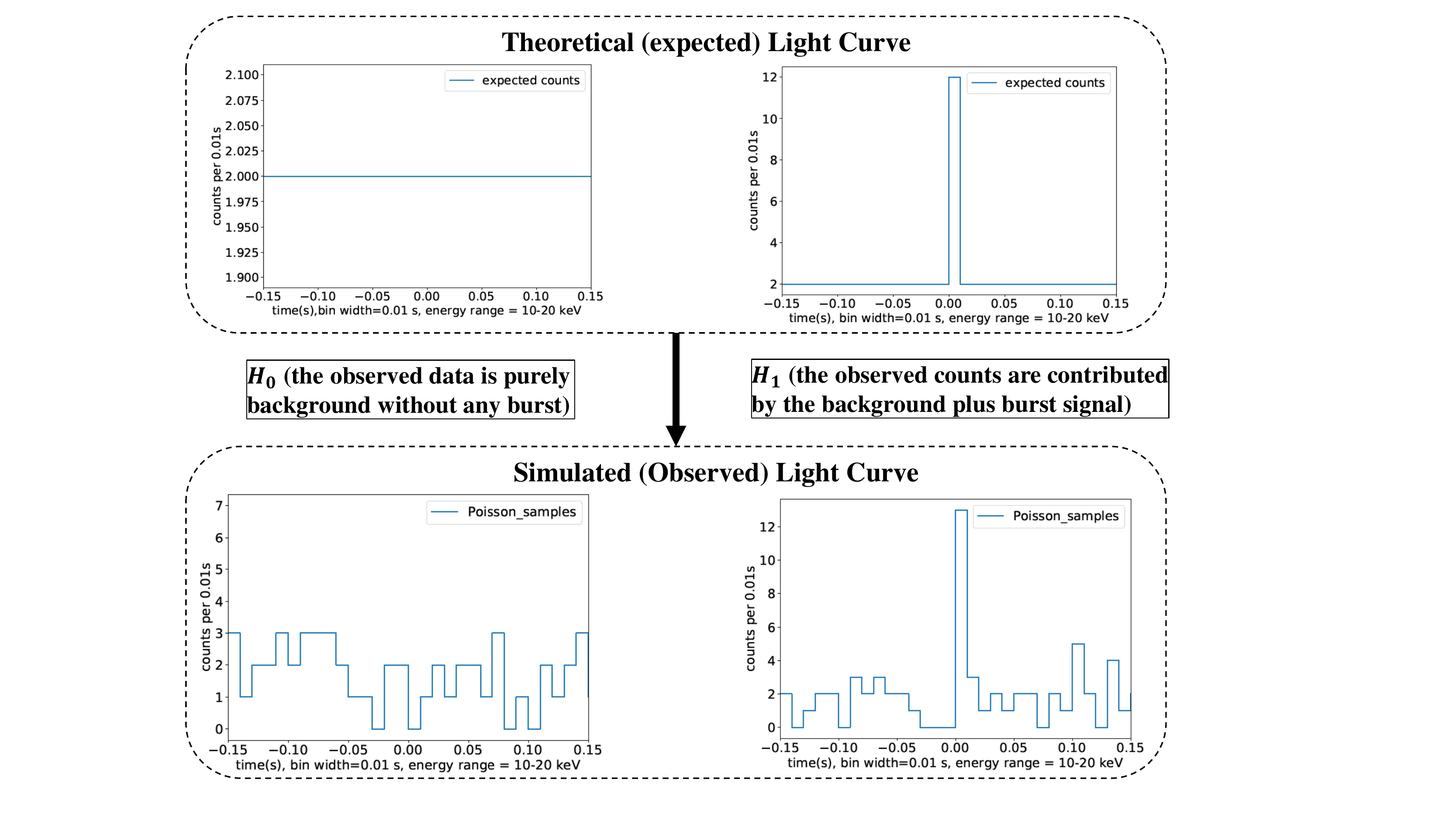}
    \caption{Theoretical Light curves for two hypothesis: the observed data is purely background without any burst signal (left), the observed counts are contributed by the background plus burst signal (right). 
    Simulated (observed) light curves are the data samples from the Poisson distribution of which the mean value is the expected counts of the theoretical light curves.}
    \label{fig:simulate_chart}
\end{figure*}

\section{Simulations and Results} \label{sec:simulation and result}

We implement a series of Monte Carlo simulations to investigate the difference between Poisson-based and Gaussian-based search methods presented in section \ref{Gaussian data} and section \ref{sec:Poisson}.  
Artificial data sets are created by the following steps:
\begin{enumerate}
    \item[(1)] Initialize the mean value of the background (i.e., $\hat{n}_{i}$) of each detector and each energy channel (i.e., $\sim$200 counts/s in 10--20 keV, $\sim$160 counts/s in 20--50 keV, $\sim$ 70 counts/s in 50--100 keV, $\sim$75 counts/s in 100-200 keV) using the GECAM data \citep{2022MNRAS.tmp..994X}.
    \item[(2)] Assume there is no evolution of the mean value of the background over time within a short time window. The known (preset) mean value of the background can be used instead of the estimated background from a polynomial fit to simulated data \footnote{In real observations, the mean value of the background is unknown, and one should use the estimated background and it's uncertainty. PGSTAT (Poisson data with Gaussian background, \citep{Arnaud2022} should be utilized in the process of searching the real observation data for burst signal. We note that performing the background estimation with PGSTAT statistic does not alter any of our main conclusions.}.  
    \item[(3)] Set the incident direction, flux amplitude, spectra shape and duration of burst source to known (simulated) values.
    \item[(4)] Calculate the burst signal by multiplying the detector response matrix of GECAM with the burst spectra  \citep[e.g. Band Function,][]{Band:1993}. 
     \item[(5)] Sample the data from Poisson distribution with $\lambda$ (expectation) equal to the mean value of the background (or the mean value of the background plus the burst signal).
\end{enumerate}

The simulated data set is created by adding the burst signal to background, as shown in Figure \ref{fig:simulate_flowchart}.
In order to assess the difference between Poisson-based and Gaussian-based search method, two different types of light curves have been studied by setting the number of data sets (i.e. detector and channel) to 1 or 100, as described in the next two subsections.

\subsection{The case for one detector and 
one channel} \label{simu_1}

For the first set of simulations, we set the number of data sets (detector and channel) to 1 (i.e., $j=1$) for simplicity\footnote{It is possible that there is only 1 detector that has good observation for some very weak bursts.}. To mimic the case for very weak and short bursts (e.g., burst duration of about 10 ms), the mean value of background is set to 2 counts (i.e., $\hat{n}_{i}=2$ in the GECAM energy range of 10--20 keV). The parameter of mean value of background is frozen in the following simulations of this subsection, unless otherwise stated.

We simulated the burst signal (i.e., $r_{i}s$) ranging from 1 counts to 24 counts. The observed counts (i.e., $d_{i}$) is background plus burst signal, and then the likelihood ratio (LR) is calculated using the known values of observed data counts, mean value of background, un-normalized signal and source amplitude($d_{i}$, $\hat{n}_{i}$, $r_{i}$, $s$).
The LR as a function of the observed counts for the Poisson and Gaussian case is shown in the left panel of Figure \ref{fig:LR_counts}, which shows that the LR in Gaussian statistics ($\mathcal L_{\rm g}$) is significantly larger than that of Poisson case ($\mathcal L_{\rm p}$) for large number of counts. This trend is opposite only for those cases with less than about 5 counts. We take the observed data of 2 counts and 20 counts as examples to show the difference of likelihood $P(d|H_{1}(s))$ and $P(d|H_{0})$ (see section \ref{sec:method} for more details of theses two likelihoods) between Poisson and Gaussian cases. As shown in the right panel of Figure \ref{fig:LR_counts}, for high observed counts (e.g., 20 counts) and low background (e.g., 2 counts), $P(d|H_{0})$ is different between Poisson and Gaussian cases, while $P(d|H_{1}(s))$ is similar to each other. For low observed counts (e.g., 3 counts) and low background (e.g., 2 counts), both $P(d|H_{0})$ and $P(d|H_{1}(s))$ show some 
difference between Poisson and Gaussian cases.

As mentioned above, we assume that there is no evolution of the mean value of the background within a short time window (for a short burst). Therefore, we focus on the time bin of signal (e.g., $T_0\sim$  $T_0+0.01 $ s) during simulation, as shown in the light curve of the right panel of Figure \ref{fig:simulate_chart}. The coherent search method of Gaussian case and Poisson case then are used to search for burst in the simulated time bin.

Assuming a source location and burst spectrum, the remaining free parameter of the likelihood ratio is burst amplitude $s$. The likelihood ratio can be maximized by choosing a proper amplitude. Therefore, the maximum likelihood ratio corresponds to the best estimation of amplitude, which is considered to be the nearest (best) value of the true source amplitude.
We test the consistency of the estimation of burst amplitude for these two search methods. The initialized mean value of signal is set to 9 counts. We run a Monte Carlo simulation and obtain a series of Poisson distributed data samples. These data samples, as unknown-amplitude observed data from several observations in the time bin of signal \footnote{Samples from Poisson distribution with expectation equal to the mean value of background (2 counts) plus signal (9 counts).}, are used to calculate the best amplitude with Newton's method that maximize $\mathcal L_{\rm g}$ and $\mathcal L_{\rm p}$, respectively. 

\begin{figure}
	\includegraphics[width=\columnwidth]{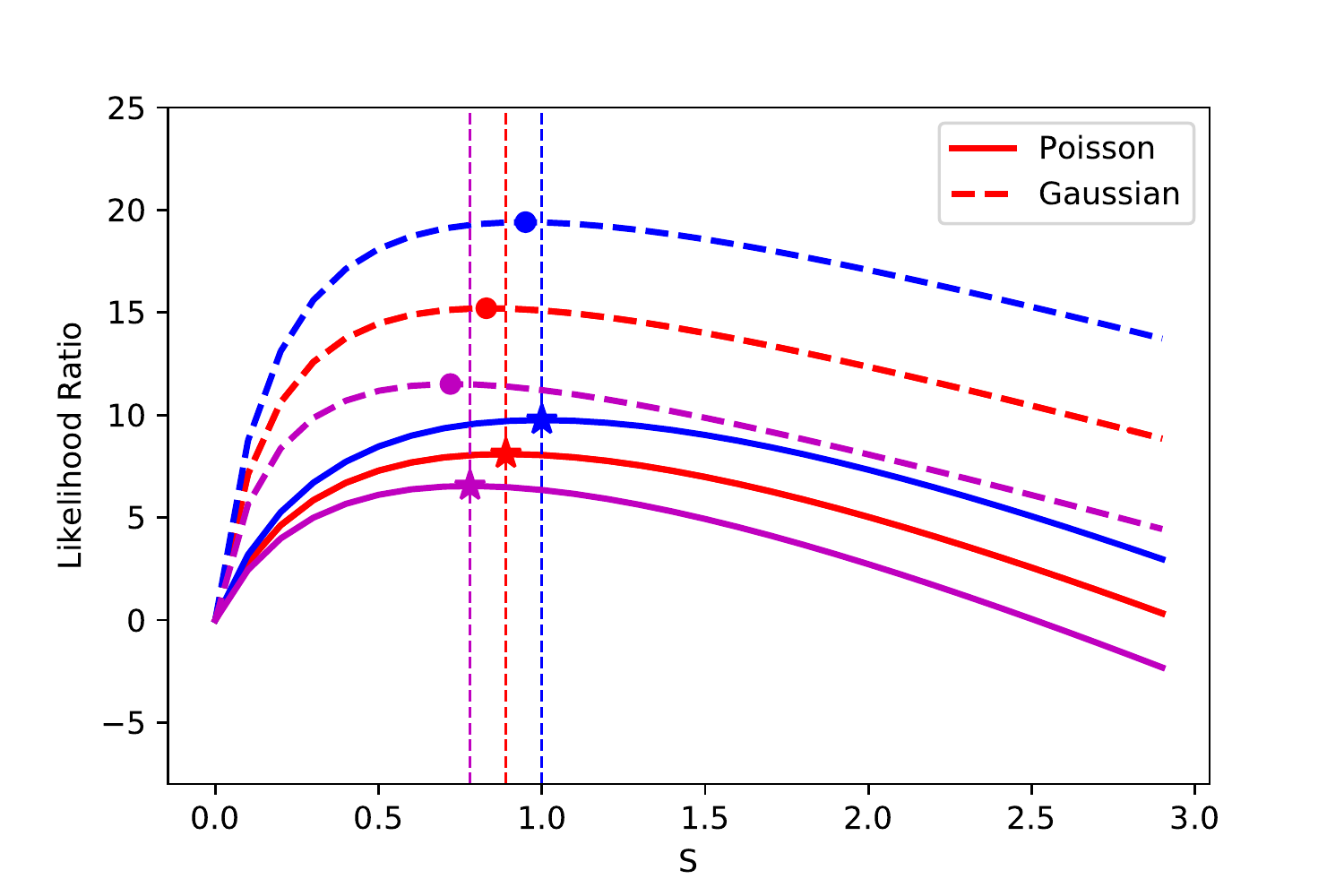}
    \caption{Likelihood ratio based on Poisson statistics (solid lines) and Gaussian statistics (dotted lines) versus source amplitude ($s$). The different colors represent different observed data samples with the same mean values of background (2 counts) and un-normalized signal (9 counts): the purple, red and blue lines are the observed data of 9 counts, 10 counts and 11 counts, respectively.} The points and stars in each dotted and solid line are the best amplitudes for Gaussian statistics and Poisson statistics that maximise likelihood ratio, respectively. The vertical lines are the observed (true) amplitude calculated using Eq.\ref{s_true} for each data sample.
    \label{fig:Sbest_value}
\end{figure}

We take three data groups (see Table \ref{tab:table_sbest} for details) as examples to show the $\mathcal L_{\rm g}$ and $\mathcal L_{\rm p}$ variation as a function of unknown amplitude of $s$. As shown in Figure \ref{fig:Sbest_value}, for each data sample, the best values of $s$ to reach the maximum of $\mathcal L_{\rm g}$ and $\mathcal L_{\rm p}$ are different for Poisson and Gaussian cases. The true amplitude of one data sample can be calculated using the known values of the mean values of background, un-normalized signal and observed data ($\hat{n}$, $r$, $d$):
\begin{equation}
{s_{\rm true}}= \frac{d-\hat{n}}{r}. \label{s_true}
\end{equation}

\begin{table*}
	\centering
	\caption{The result of estimation of best amplitude and likelihood ratio in Gaussian statistics and Poisson statistics.}
	\begin{threeparttable}
	\label{tab:table_sbest}
	\begin{tabular}{cccccccccc} 
		\hline
		\hline
		number & $\hat{n}$ & $r$ & $d$  &  $S_{\rm Poisson}$ & $S_{\rm Gaussian}$ & $\mathcal L_{\rm g}(S_{\rm Gaussian})$ & $\mathcal L_{\rm g}(S_{\rm Poisson})$ & $\mathcal L_{\rm p}(S_{\rm Gaussian})$ & $\mathcal L_{\rm p}(S_{\rm Poisson})$\\
		\hline
		1 & 2 & 9 & 10 & 0.89 & 0.83 & 15.20 & 15.19 & 8.07 & 8.09\\
		2 & 2 & 9 & 11 & 1.00 & 0.95 & 19.40 & 19.39 & 9.74 & 9.75\\
		3 & 2 & 9 & 9 & 0.78 & 0.72 & 11.51 & 11.49 & 6.52 & 6.53 \\
		\hline
	\end{tabular}
    \begin{tablenotes}
     \footnotesize
     \item[1] $\hat{n}$ and $r$ are the mean values of background and un-normalized signal with the initialized source amplitude of 1, respectively. 
     \item[2] $d$ is the observed data counts arise from the Poisson process with the mean value of $\hat{n}+r$.
     \item[3] $S_{\rm Poisson}$ and $S_{\rm Gaussian}$ are the estimation of best source amplitude given by the  Poisson case and Gaussian case, respectively.
     \item[4] $\mathcal L_{\rm g}(S_{\rm Gaussian})$ and $\mathcal L_{\rm g}(S_{\rm Poisson})$ are Gaussian statistics based likelihood ratio with  $S_{\rm Gaussian}$ and $S_{\rm Poisson}$, respectively.
     \item[5] $\mathcal L_{\rm p}(S_{\rm Gaussian})$ and $\mathcal L_{\rm p}(S_{\rm Poisson})$ are Poisson statistics based likelihood ratio with $S_{\rm Gaussian}$ and $S_{\rm Poisson}$, respectively.
    \end{tablenotes}
    \end{threeparttable}
\end{table*}

As a test example, for the case of $\hat{n}=2$, $r=9$ and $d=10$ (see the red line of Figure \ref{fig:Sbest_value} and Table \ref{tab:table_sbest}), $s_{\rm true}$ equals to 0.89, which is consistent with the estimation of amplitude ($S_{\rm Poisson}$, 0.89) when maximize the likelihood ratio using the Poisson case. The value of $s_{\rm true}$ is a little different from the estimation of amplitude ($S_{\rm Gaussian}$, 0.83) when maximize the likelihood ratio using the Gaussian case. Given the values of $S_{\rm Poisson}$ and $S_{\rm Gaussian}$, we can calculate the likelihood ratio based on Gaussian case ($\mathcal L_{\rm g}(S_{\rm Gaussian})$ and $\mathcal L_{\rm g}(S_{\rm Poisson})$) using Eq. \ref{eq3} and based on Poisson case ($\mathcal L_{\rm p}(S_{\rm Gaussian})$ and $\mathcal L_{\rm p}(S_{\rm Poisson})$) using Eq. \ref{eq4}, respectively. We note that $\mathcal L_{\rm g}(S_{\rm Gaussian})$ is larger than $\mathcal L_{\rm g}(S_{\rm Poisson})$ and $\mathcal L_{\rm p}(S_{\rm Poisson})$ is larger than $\mathcal L_{\rm p}(S_{\rm Gaussian})$, which shows that $S_{\rm Gaussian}$ is better than $S_{\rm Poisson}$ for Gaussian statistics and $S_{\rm Poisson}$ is better than $S_{\rm Gaussian}$ for Poisson statistics.
The results of best amplitude and maximum LR from all three data groups are listed in Table \ref{tab:table_sbest}. 

It is usually not very reliable to claim the detection of a burst when there is only a small excess in one detector and one channel, so more detailed simulations and analyses with detection of multi-detectors and multi-channels are performed in next subsection.

\subsection{The case for multiple detectors and channels}

\begin{figure}
    \includegraphics[width=0.5\textwidth]{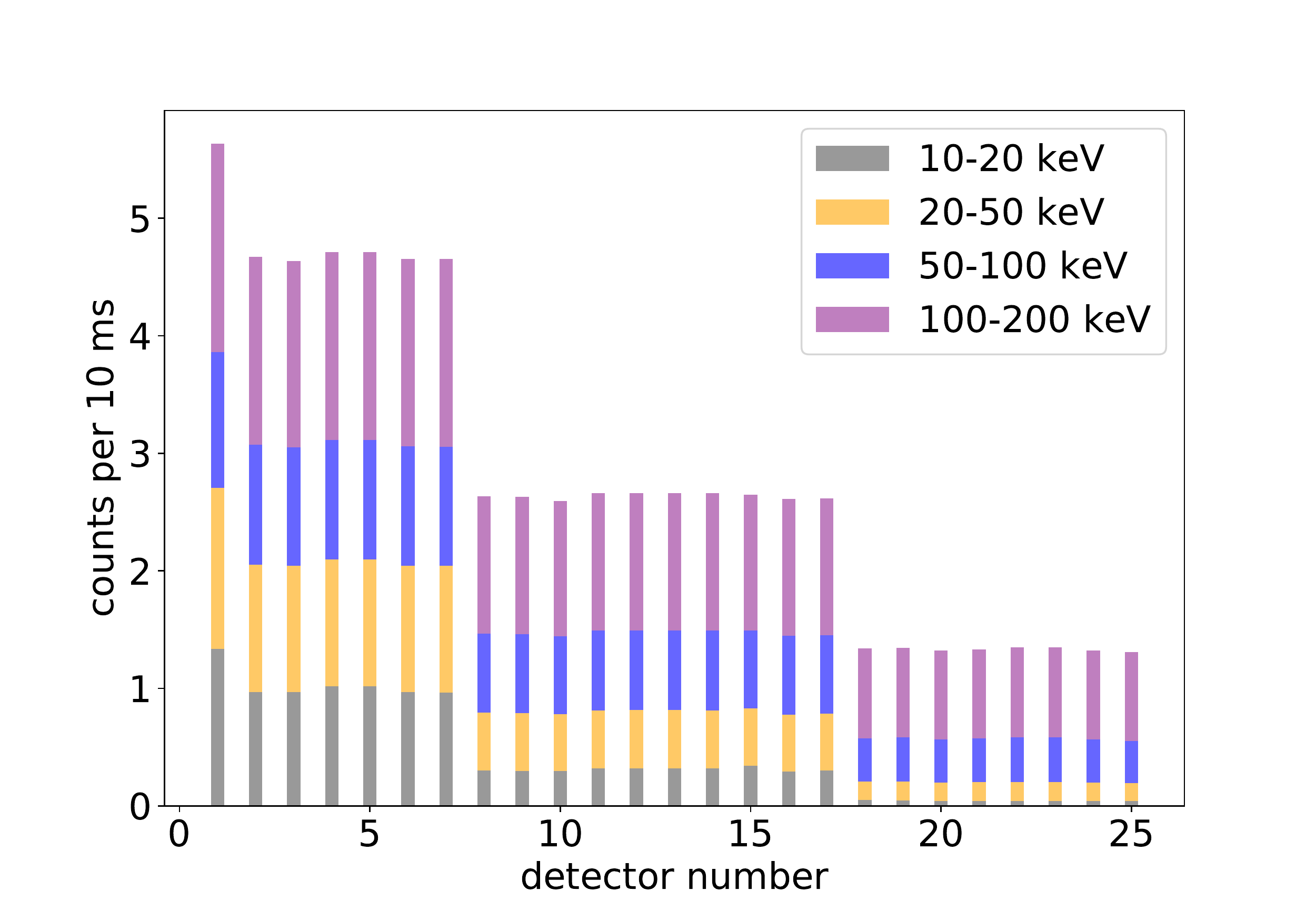}
    \caption{The expected counts of all 25 GRD detectors of GECAM for simulated burst is calculated using the assumed location (i.e., theta = 0 and phi = 0 in payload coordinate system), the power law spectrum and the detector response matrix of GECAM \citep{QiaoRui2022}. Different colors represent different energy bands. The gray color is the energy band of 10--20 keV; the orange color is the energy band of 50--100 keV; the blue color is the enegy band of 50--100 keV; the purple color is the energy band of 100--200 keV.}
    \label{fig:simulate_counts}
\end{figure}

\begin{figure*}
    \begin{tabular}{cc}
	\includegraphics[width=\columnwidth]{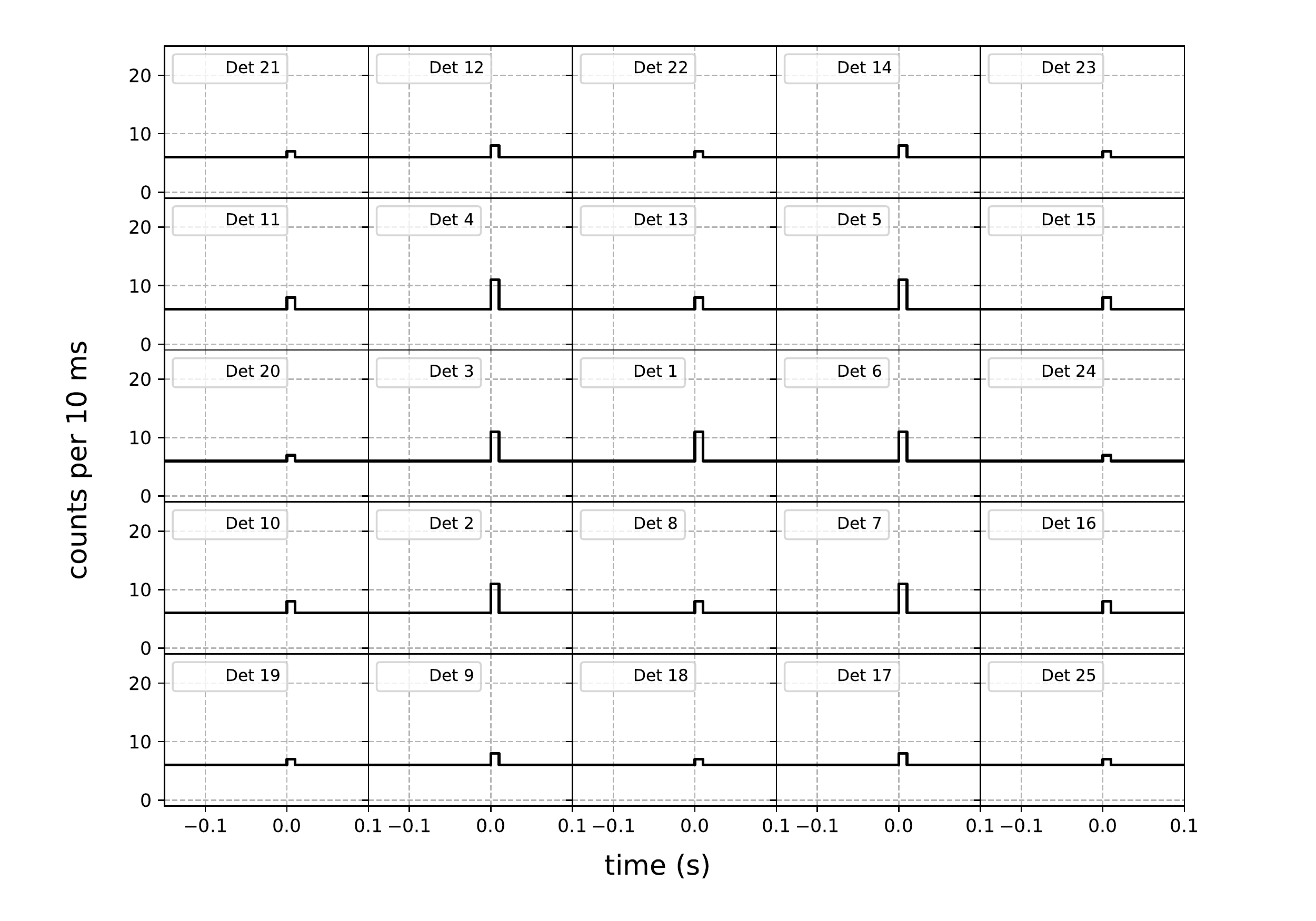} &
	\includegraphics[width=\columnwidth]{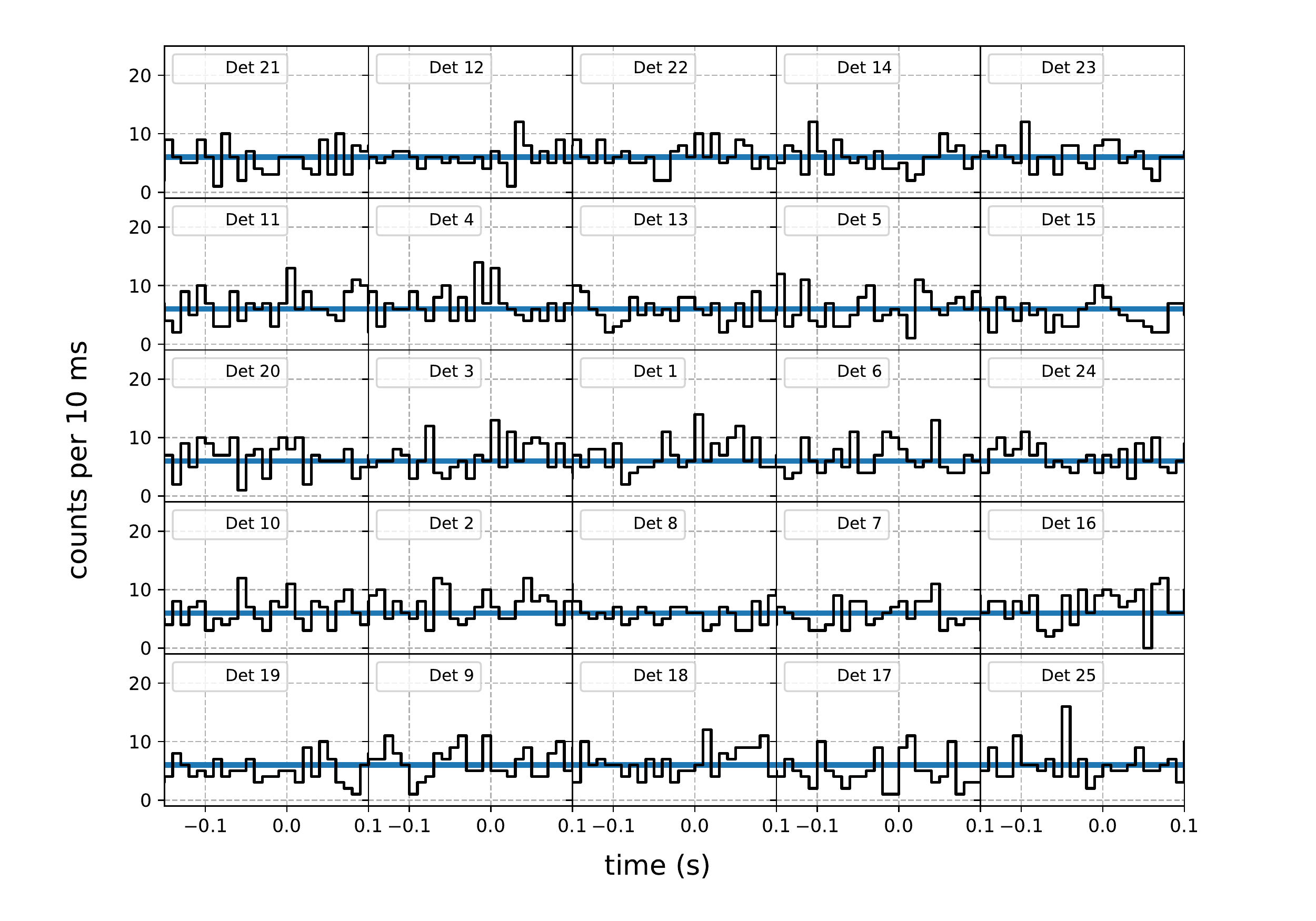} \\
	\end{tabular}
	\caption{GECAM Light curves of simulated burst in the energy range of 10--200 keV. $T_{\rm 0}$ is the start time of this burst. \textit{Left}: Theoretical light curves of each detector. These counts is equal to the mean value of background plus the burst expected counts. \textit{Right}: Observed (simulated) light curves of each detector. These counts is sampled from Poisson distribution with the expectation equal to the counts of the theoretical light curves in each time bin. The blue lines represent the estimated background (i.e., the mean value of the background). 
    }
    \label{fig:mulitlc1}
\end{figure*}

The most probable application of the coherent search method is the detection of weak burst sources with multi-detectors and multi-channels.
We simulate the burst signal detected by 25 detectors and 4 channels (for each detector) of GECAM using the response matrix in previous work \citep{QiaoRui2022}. For simplification, we assume that the source has constant flux during burst time and the background also stay stable within short time windows. All 25 detectors have the same background level.

As mentioned in section \ref{introducion}, it is important to find and study weak and short bursts. We set the duration of the burst (e.g., magnetar bursts) to 10 ms \citep{2021GCN.30140....1C}. Typically the fluences of magnetar bursts detected by GBM (or GECAM) mostly range from $\sim 10^{-8}$ erg cm$^{-2}$ to $\sim 10^{-6}$ erg cm$^{-2}$ in the energy range of 10 keV -- 200 keV \citep{2015ApJS..218...11C,XiongShaoLin2022}. We set the spectrum and flux of the weak short burst in this simulation to the Power Law function\footnote{A simple photon power law with photon index of $\alpha$ and normalization of $S$ (in units of photons cm$^{-2}$ s$^{-1}$ keV$^{-1}$): ${f}_{\rm PL}(E) = S \times {(\frac{E}{100})}^{-\alpha}$} and $1\times 10^{-8}$ erg cm$^{-2}$ s$^{-1}$(10--200 keV). The expected counts of each GECAM detector for simulated burst are shown in Figure \ref{fig:simulate_counts}, which are added to the background to simulate the observed light curve. The maximum and minimum number of expected burst counts of each detector and channel are about $\sim$2 counts and $\sim$0 counts. The total number of expected burst counts and background are 63 counts and 150 counts, respectively.

\begin{figure}
    \begin{tabular}{c}
	\includegraphics[width=\columnwidth]{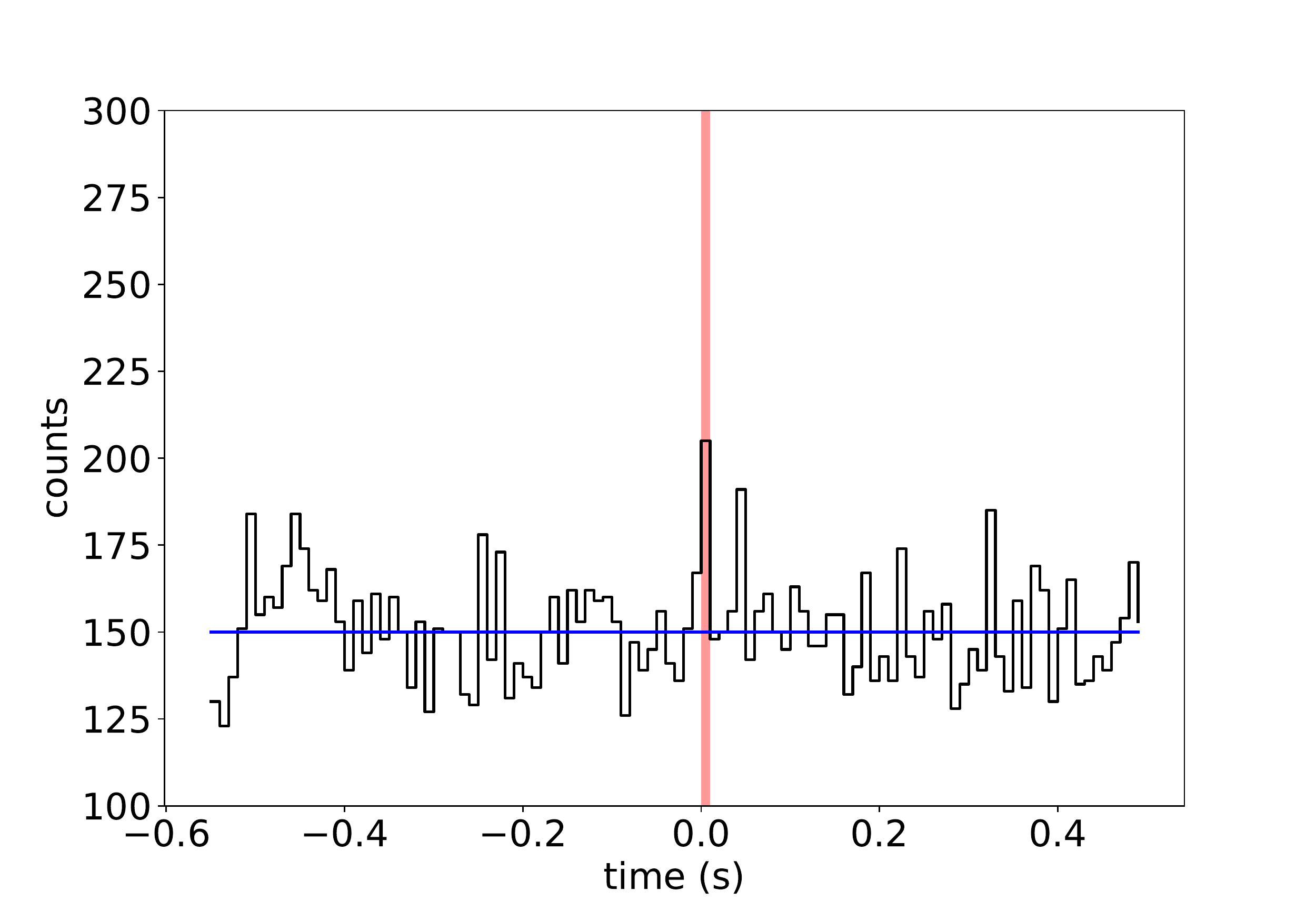}
	\end{tabular}
	\caption{Summed light curve of all 4 channels and 25 detectors for a simulated weak burst. The blue line is estimated background. $T_{\rm 0}$ is the start time of this burst. The bursts with shorter duration (10 ms) is marked by red bar.}
	\label{fig:mulitlc_2}
\end{figure}

The theoretical light curves of each GECAM detector for the weak burst are shown in the left panel of Figure \ref{fig:mulitlc1}.  The simulated light curves are shown in the right panel of Figure \ref{fig:mulitlc1}, which are sampled from Poisson distribution with the expected values equal to the counts in each time bin of the theoretical light curves. There are some signals with very few counts (less than about 5 counts) of each channel and detector for such kind of short weak bursts. 
All counts of each channel and detector unit are summed together for this weak burst, as shown in Figure \ref{fig:mulitlc_2}. It is shown that there are very limited excess counts in the observed light curve for such weak short burst.

\begin{figure}
    \begin{tabular}{cc}
	\includegraphics[width=\columnwidth]{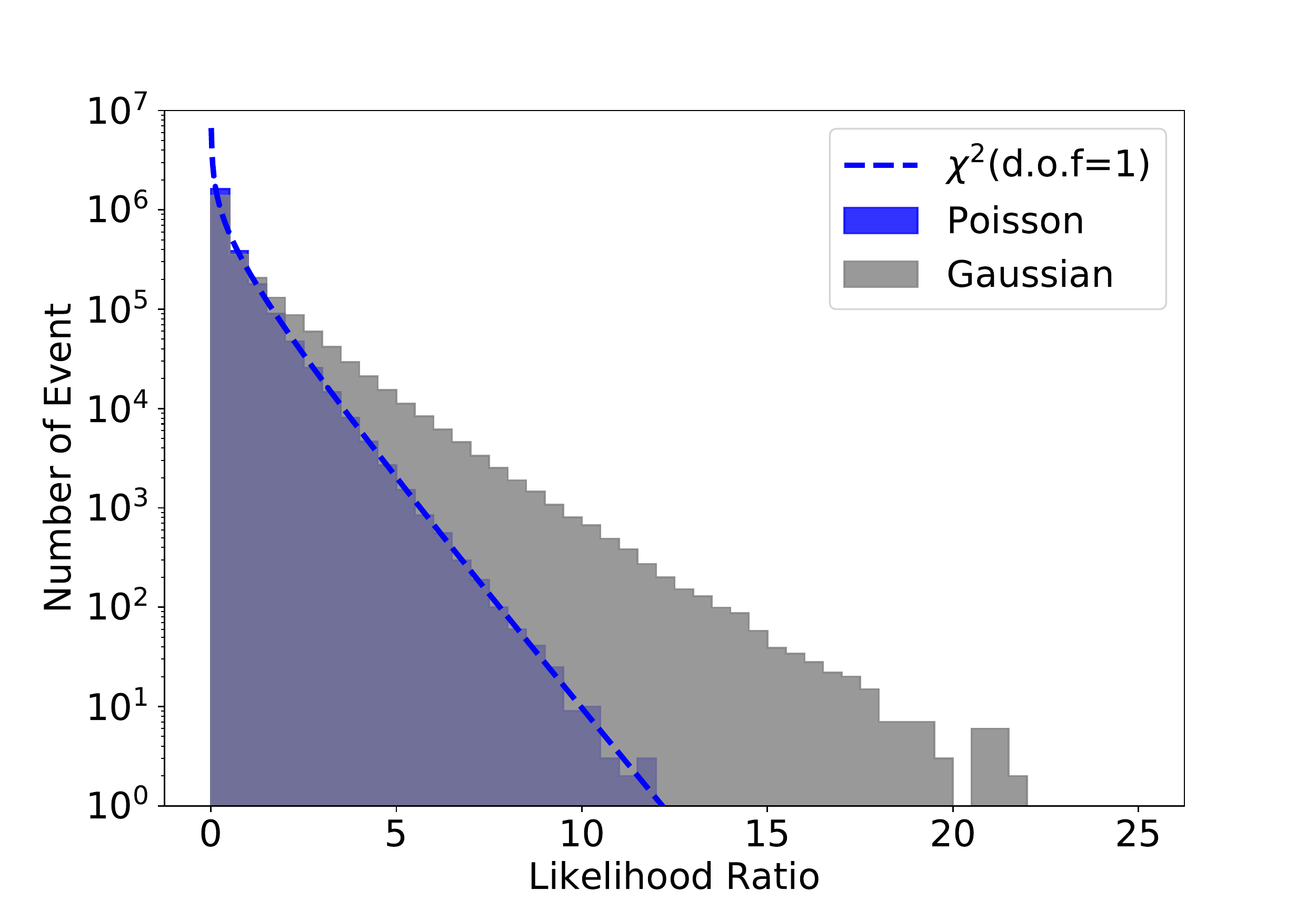} \\
	\end{tabular}
	\caption{Simulated distributions for likelihood ratio using background alone. The gray and blue color represent Gaussian case and Poisson case, respectively. The value of $2\mathcal L$ (equal to the form of likelihood ratio in Wilks's theorem) tends to a $\chi^{2}$ distribution with the freedom of 1 (blue dotted line). The mean values of the background are 2 counts, 2 counts, 1 counts and 1 counts with the duration of 10 ms and the energy ranges of 10--20 keV, 20--50 keV, 50--100 keV, 100--200 keV, which are based on the background level of GECAM. 
	}
	\label{fig:Dist_Bakc_LR}
\end{figure}

\begin{figure*}
    \begin{tabular}{cc}
    \includegraphics[width=\columnwidth]{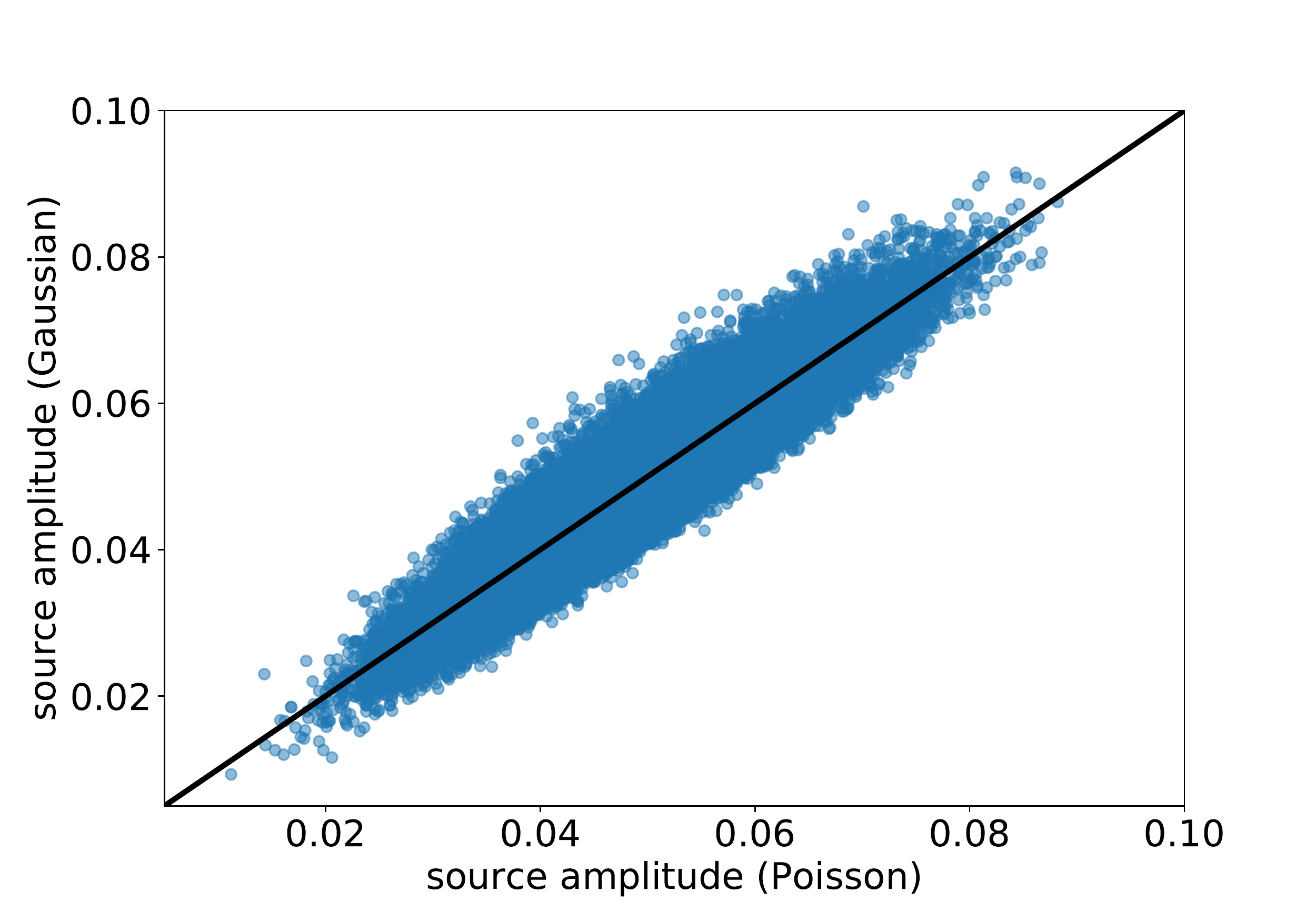} &
	\includegraphics[width=\columnwidth]{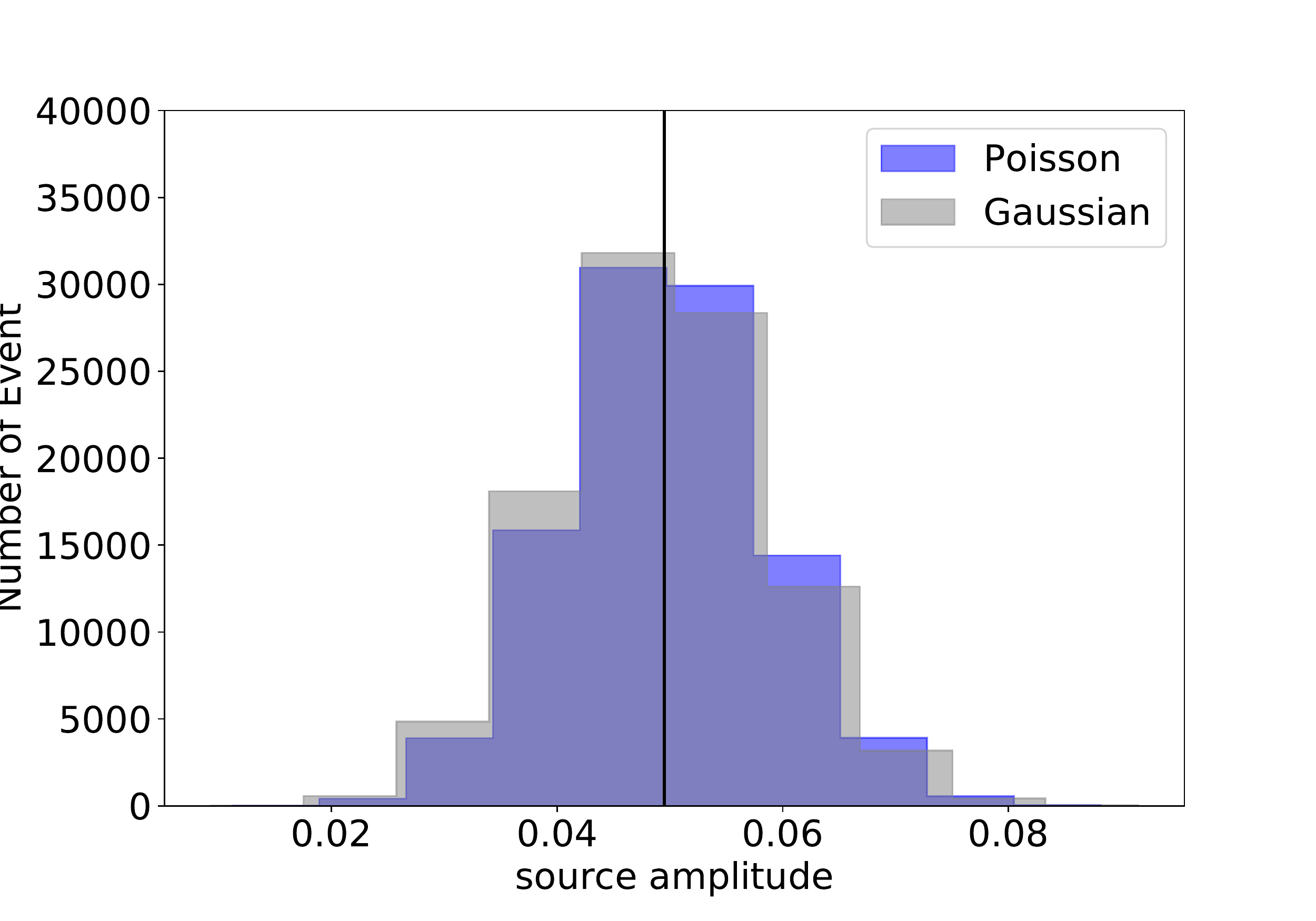} \\
	\end{tabular}
	\caption{\textit{Left}: Scatter plot of the source amplitude from Gaussian case and Poisson case. The black line represents the equivalence between these two source amplitudes. \textit{Right}: The distribution of source amplitude. The gray and blue color represent the Gaussian case and Poisson case, respectively.
	The black line represents the expected amplitude.}
	\label{fig:Dist_s}
\end{figure*}

The coherent search methods based on Poisson statistics and Gaussian statistics are used to search for this weak burst in the simulated light curve. We calculate the likelihood ratio ($\mathcal L_{\rm g}$ and $\mathcal L_{\rm p}$) of the time bin of the burst signal, using the values of estimated background and un-normalized signal (with the known location and spectrum of the source preset). The likelihood ratio can be maximized by varying the amplitude. This results in the best amplitude of 0.03 for both the Poisson case and Gaussian case, which has a little different from the preset amplitude (0.04) due to the Poisson fluctuation in the simulation. 
The maximum likelihood ratios corresponding to the best amplitude are 11.17 and 16.29 for Poisson case and Gaussian case, respectively. The resulted absolute values of $\mathcal L_{\rm g}$ and $\mathcal L_{\rm p}$ are different, which is the same as the case for one detector and one channel (see section \ref{simu_1}).

We calibrate the corresponding confidence level for these two specific likelihood ratios by simulating the procedure of searching for signals in the simulated light curve including only Poisson noise. Using the same background level of the simulated weak burst, we executed a series of simulations to investigate the LR distribution for both Poisson and Gaussian cases, as shown in Figure \ref{fig:Dist_Bakc_LR}. 
We find that the LR of Poisson case ($2\mathcal L_{\rm p}$) follows the $\chi^2$ distribution with the degree of freedom (d.o.f.) of 1, even for the extremely low counts. By contrast, the LR of the Gaussian case ($2\mathcal L_{\rm g}$) significantly deviates from this $\chi^2$ distribution in low counts regime.

With this LR distribution, we can estimate the significance of a specific likelihood ratio (T) according to: 
\begin{equation}
P_{\rm T} = \frac{N_{\rm LR>T}}{N_{\rm {total}}}, \label{sig}
\end{equation}
where $N_{\rm LR>T}$ and $N_{\rm {total}}$ are the number of simulated events with LR larger than T and total number of simulation events, respectively.

For this simulated burst mentioned above, the LR of Poisson case (11.17) and Gaussian case (16.29) correspond to the confidence level (calculated using Eq. \ref{sig}) of $2.51\times10^{-6}$ and $5.32\times10^{-5}$, respectively.

We also run a series of simulations to assess the estimation of source amplitude and confidence level for Gaussian case and Poisson case with the same expected burst counts and background. 
Given the values of background, un-normalized signal and observed counts, we can calculate the best amplitudes corresponding to maximum $\mathcal L_{\rm g}$ and $\mathcal L_{\rm p}$.

The source amplitude of Gaussian case is correlated with that of Poisson case, as shown in the left panel of Figure \ref{fig:Dist_s}.  The distributions of the source amplitude of Gaussian case and Poisson case are shown in the right panel of Figure \ref{fig:Dist_s}. We find that there are no significant differences between the amplitude of Gaussian case and Poisson case.

Based on the result of the simulations with background only, the significance of these specific likelihood ratios ($P_{\rm Gaussian}(\mathcal L_{\rm g})$ and $P_{\rm Poisson}(\mathcal L_{\rm p})$) can be calculated using Eq. \ref{sig}.
These detailed results mentioned above as well as the significance given by $\chi^2$ distribution at 2$\mathcal L_{\rm p}$ (equal to the form of the likelihood ratio in Wilk's theorem)
are listed in Table \ref{tab:Prob}. Interestingly, we find that, the significance of a burst given by Gaussian case is lower than that of the Poisson case.

\begin{table*}
    \centering
    \caption{Test result for significance in Gaussian statistics and Poisson statistics.}
    \begin{threeparttable}
    \label{tab:Prob}
    \begin{tabular}{cccccccc} 
        \hline
        \hline
         number & $S_{\rm Poisson}$ & $S_{\rm Gaussian}$ &$\mathcal L_{\rm p}$ & $\mathcal L_{\rm g}$ & $P^a_{\rm Poisson}(\mathcal L_{\rm p})$& $P^b_{\rm Gaussian}(\mathcal L_{\rm g})$  &   $P^c_{\chi^2}(2\mathcal L_{\rm p})$\\
        \hline
         1 & 0.030 & 0.030 & 11.17 & 16.29  &$2.51\times 10^{-6}$  & $5.32\times 10^{-5}$ & $2.28\times 10^{-6}$\\
         2 & 0.031 & 0.031 & 9.47 & 15.62 &$2.22\times 10^{-5}$  & $6.74\times 10^{-5}$ & $1.34\times 10^{-5}$\\
         3 & 0.032 & 0.036 & 10.13 & 18.47 & $7.96\times 10^{-6}$ & $1.76\times 10^{-5}$ & $6.75\times 10^{-6}$\\
         4 & 0.035 & 0.034 & 12.17 & 17.51 &$4.19\times 10^{-7}$ &  $2.38\times 10^{-5}$ & $8.07\times 10^{-7}$\\
         5 & 0.030 &  0.030 & 8.79 & 13.86 &$3.93\times 10^{-5}$ &  $1.86\times 10^{-4}$ & $2.75\times 10^{-5}$\\
        \hline
    \end{tabular}
    \begin{tablenotes}
     \footnotesize
     \item[a,b] The significance in Poisson case and Gaussian case at $\mathcal L_{\rm p}$ and $\mathcal L_{\rm g}$, respectively. 
     \item[c] The significance given by $\chi^2_1$ distribution at $2\mathcal L_{\rm p}$ (equal to the form of likelihood ratio in Wilks's theorem). 
    \end{tablenotes}
    \end{threeparttable}
\end{table*}

\begin{figure*}
    \begin{tabular}{cc}
	\includegraphics[width=\columnwidth]{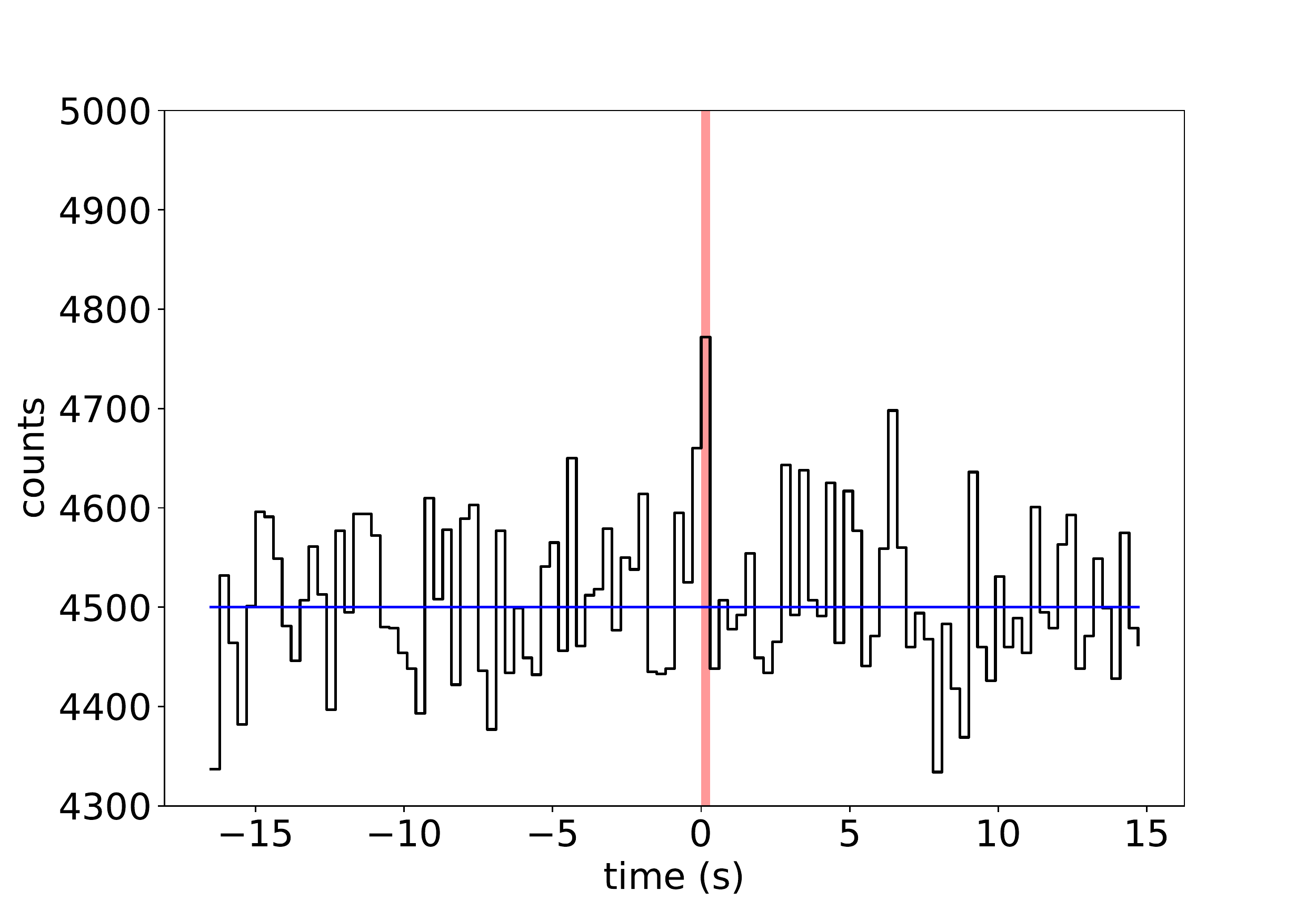} &
	\includegraphics[width=\columnwidth]{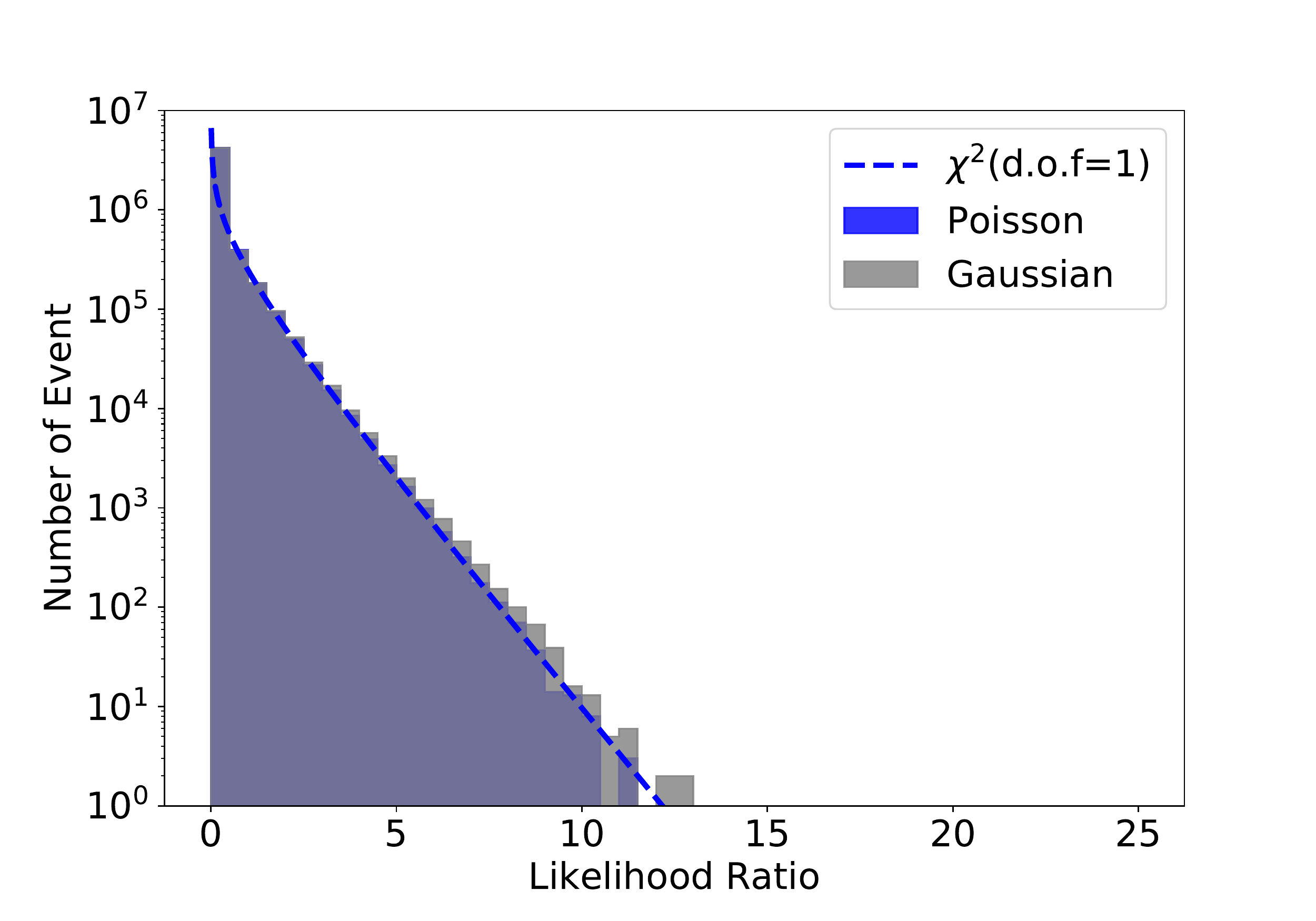} \\
	\end{tabular}
	\caption{\textit{Left}: Summed light curve of all 4 channels and 25 detectors for a simulated weak burst with high level of background. 
	The blue line is estimated background. $T_{\rm 0}$ is the start time of this burst. The bursts with duration of 300 ms, is marked by vertical red bar. 
	The mean values of the background are 60 counts, 60 counts, 30 counts and 30 counts with the duration of 300 ms and the energy range of 10--20 keV, 20--50 keV, 50--100 keV, 100--200 keV, which are based on the background level of GECAM.
	\textit{Right}: Simulated distributions for likelihood ratio using the same background counts level as the light curve of left panel. Other captions are same as Figure \ref{fig:Dist_Bakc_LR}}.
	\label{fig:mulitlc_3}
\end{figure*}

High counts levels are also used to study the difference of LR between the Poisson case and Gaussian case. The simulated light curve of each detector and each channel are sampled from Poisson distribution with the expected value of high counts (more than 20 counts). We assume that the duration of the simulated burst is 300 ms. The mean values of the background (i.e, 60 counts, 60 counts, 30 counts, 30 counts) for each energy channel (i.e., 10--20 keV, 20--50 keV, 50--100 keV and 100--200 keV) correspond to the in-flight background level of GECAM with the duration of 300 ms. The total number of expected signal counts and background are set to 282 and 4500. Summed light curve of all 4 channels and 25 detectors for the simulated weak burst are shown in the left panel of Figure \ref{fig:mulitlc_3}. 
For the time bin of this burst, the estimated best amplitude and the corresponding likelihood ratio is 0.05 and 29.14 for Poisson case, which is well consistent with the estimated amplitude and likelihood ratio of 0.05 and 31.64 for Gaussian case. We note that this result for high counts is different from the case of low counts (i.e., low level of background and signal) mentioned above. The corresponding confidence level for the case of high counts level are also calibrated. The LR distribution of pure background variation for Poisson case and Gaussian case are shown in the right panel of Figure \ref{fig:mulitlc_3}. It shows that the LR distribution for these two statistics are generally agreement with each other and both of them follow the $\chi^2$ distribution with the degree of freedom (d.o.f.) of 1.

\section{Discussions} \label{sec:Discussions}

In this paper, we proposed the coherent search based on the Poisson statistics to search for all kinds of bursts, especially for weak and short bursts. We implemented a series of simulations (see Figure \ref{fig:simulate_flowchart} for simulation framework) to evaluate the difference between the coherent search methods based on Poisson statistics and Gaussian statistics. 

Two different number of data sets (i.e., $j$) are used for simulations: one detector and one channel, twenty-five detectors and four channels. For the case of one detector and one channel, the absolute value of LR and estimation of amplitude can be simply and directly compared with each other, which is very important for understanding the intrinsic difference between these two statistics.
However, it is more common to search out a burst with multi-detectors and multi-channels (e.g., 25 detectors and 4 channels). Therefore, multi-detectors and multi-channels based on GECAM data have been implemented in our simulations.

As mentioned in section \ref{sec:method}, the LR based on Poisson and Gaussian distribution may show significant difference for those two cases when the mean value of counts is small and the number of variables that contribute to the sum of likelihood in not large, i.e. the Gaussian distribution is not a good approximation any more. Therefore, we focus on the low counts level of background for weak and short bursts. Furthermore, high counts level of background is also applied to validate the consistency of these two statistics.

\subsection{Comparison of the absolute values of LR}

For the case of one detector and one channel, the likelihood ratio based on Poisson statistics is different from that of the Gaussian statistics in the low level of background. To avoid the error of statistical fluctuations, the simulated observation data is set to the mean value of background plus the mean value of signal. As shown in the left panel of Figure \ref{fig:LR_counts}, for low background level (during a short duration of the burst, e.g., 10 ms), the LR is different both at low observed counts (e.g., weak bursts) and high observed counts (e.g., bright bursts). Since the LR is calculated using two probabilities ($P(d|H_{1}(s)$, $P(d|H_{0})$) with different mean values, the differences of LR is the reflection of these two probabilities. The distributions of Poisson and Gaussian with different mean values in the linear scale and logarithm scale are shown in the right panel of Figure \ref{fig:LR_counts}, which demonstrates that $P(d|H_{0})$ of Poisson case at low expected counts (e.g., the background level of 2 counts) and high observed counts (i.e., background plus signal) are very different from that of Gaussian case. With a fewer observed counts, $P(d|H_{1}(s))$ is also different between these two distributions. We also find that both $P(d|H_{1}(s))$ and $P(d|H_{0})$ are similar to each other at high expected background counts (e.g., the background level of 20 counts). 

We compared the results of low counts region with different number of data sets (i.e. 25 detectors and 4 channels) and find that it does not alter any of our main conclusions about the difference of LR between Poisson case and Gaussian case. 

\subsection{Comparison of the estimation of burst amplitude}

As an important parameter for studying the brightness of bursts, the amplitude can be calculated through maximizing the LR.

For the case of one detector and one channel, the estimation of source amplitude is unbiased even for extremely weak burst using Poisson statistics. However, for Gaussian statistics, the estimation of source amplitude is slightly biased for weak burst.
The correctness and difference of the estimation of source amplitude using these two statistics are test by Monte Carlo simulations. Simulated light curves are sampled from the Poisson distribution with the mean value of theoretical light curves, as shown in Figure \ref{fig:simulate_chart}. For the case of one detector and one channel, three simulated data groups with the same mean values of background (2 counts) and un-normalized signal (9 counts) are used to compare the amplitude of Poisson case and Gaussian case. Figure \ref{fig:Sbest_value} shows that there is a little difference between the best estimated amplitudes of Poisson case and Gaussian case. We calculate the true amplitude of each simulated data sample using Eq.\ref{s_true} and find that the estimated amplitude of Poisson case is well consistent with the true value. 
The detailed results are listed in Table \ref{tab:table_sbest}, which shows that the estimated amplitudes of Poisson case and Gaussian case can maximize LR of Poisson case and Gaussian case, respectively. We note that using the estimated amplitudes of Poisson case can not maximize the LR of Gaussian case, and vice versa.

For the case of 25 detectors and 4 channels, a weak magnetar burst is simulated to compare the Poisson case and Gaussian case.
Assuming the weak bursts are detected by GECAM, it results in the maximum counts of 2 and the minimum counts of 0 in each detector and channel, as shown in Figure \ref{fig:simulate_counts}.
Figure \ref{fig:mulitlc1} and Figure \ref{fig:mulitlc_2} show the light curve of this simulated burst with low level of background and signal. We find that the best amplitude of Poisson case for this simulated burst is in agreement with the amplitude of Gaussian case. Figure \ref{fig:Dist_s} also show that there are no significant differences between the amplitude of Gaussian case and Poisson case with a series of simulations \footnote{We note that Eq.\ref{s_true} is not appropriate for the case of multi-detectors and multi-channels with the consideration of the optimized weighted light curve (see \cite{10.1093/mnras/stab2760} for more details). }.


\subsection{Comparison of the confidence level} 

Two common techniques to estimate the significance of observed burst signals are utilized in many studies: analytical calculation with formulae and direct calculation with simulations. According to Wilks's theorem, the likelihood ratio (equal to the twice of the likelihood ratio in this work) approaches to the $\chi^2_1$ distribution as the number of data samples tends to infinity. Thus, the significance can be roughly estimated with $\chi^2_1$ distribution in the case of large data samples. 

Our simulations show that, in low counts regime (e.g. less than 10 counts),
only the LR of the simulated data from background fluctuations in Poisson statistics (i.e. $2\mathcal L_{\rm p}$) generally follow the $\chi^2_1$ distribution, while the LR in Gaussian case deviate the $\chi^2_1$ distribution significantly, as shown in Figure \ref{fig:Dist_Bakc_LR}. We also find that in high counts regime, the LR of Gaussian case is well consistent with that of Poisson case, as shown in Figure \ref{fig:mulitlc_3}. 

Since the observation data are generally limited by the sensitivity of detectors, background level and the brightness of source, one must be cautious to use the Wilks's theorem to calculate the significance of a burst candidate. Here, we make use of simulations with low background level to assess the confidence level for Poisson case and Gaussian case.
For the case of multi-detectors and multi-channels, the significance of the simulated burst (e.g figure \ref{fig:mulitlc_2}) are different between Poisson statistics and Gaussian statistics. Our results show that the Poisson-based search can provide higher significance of the burst than the Gaussian-based search (see table \ref{tab:Prob}).

A weak burst with high background level is also simulated to validate the consistency between Poisson case and Gaussian case, as shown in Figure \ref{fig:mulitlc_3}. We find that the LR, estimation of source amplitude and confidence level for Poisson case are well consistent with those of Gaussian case.

\section{Conclusions} \label{sec:CONCLUSIONS}

In this paper, we proposed the coherent search based on likelihood ration with Poisson statistics. We compared the likelihood ratio values given by Poisson and Gaussian statistics, derived the best amplitude of the burst using Newton's method and presented the results of simulations for different setting of background and burst signal levels, including the results of pure background fluctuations.

 We find that the Poisson-based search method has advantages than that based on Gaussian statistics, especially for weak and short bursts. When the counts number is very low (which is usual for very short burst down to ms time scale), the Poisson-based search can provide higher significance than the Gaussian-based search and its likelihood ratio still follows the $\chi^{2}$ distribution, which provides a fast estimation of the burst significance.
 
 Although these two methods should be basically equal for bright bursts with large number of counts, to deal with general bursts including weak and short ones, we suggest that the Poisson-based coherent search (see section \ref{sec:Poisson}) should be used in transients search and study.

\section*{Acknowledgements}
This work is supported by the National Key R\&D Program of China (2021YFA0718500).
We thank supports from 
the Strategic Priority Research Program on Space Science, the Chinese Academy of Sciences (Grant No.
XDA15360102, 
XDA15360300, 
XDA15052700) 
, the National Natural Science Foundation of China (Grant No. 
12173038 
, U2038106) 
and the National HEP Data Center (Grant No. E029S2S1). 

\section*{Data Availability}
The data and codes underlying this article will be shared on reasonable request to the corresponding author.



\bibliographystyle{mnras}
\bibliography{main} 

\begin{thebibliography}{}
\makeatletter
\relax
\def\mn@urlcharsother{\let\do\@makeother \do\$\do\&\do\#\do\^\do\_\do\%\do\~}
\def\mn@doi{\begingroup\mn@urlcharsother \@ifnextchar [ {\mn@doi@}
  {\mn@doi@[]}}
\def\mn@doi@[#1]#2{\def\@tempa{#1}\ifx\@tempa\@empty \href
  {http://dx.doi.org/#2} {doi:#2}\else \href {http://dx.doi.org/#2} {#1}\fi
  \endgroup}
\def\mn@eprint#1#2{\mn@eprint@#1:#2::\@nil}
\def\mn@eprint@arXiv#1{\href {http://arxiv.org/abs/#1} {{\tt arXiv:#1}}}
\def\mn@eprint@dblp#1{\href {http://dblp.uni-trier.de/rec/bibtex/#1.xml}
  {dblp:#1}}
\def\mn@eprint@#1:#2:#3:#4\@nil{\def\@tempa {#1}\def\@tempb {#2}\def\@tempc
  {#3}\ifx \@tempc \@empty \let \@tempc \@tempb \let \@tempb \@tempa \fi \ifx
  \@tempb \@empty \def\@tempb {arXiv}\fi \@ifundefined
  {mn@eprint@\@tempb}{\@tempb:\@tempc}{\expandafter \expandafter \csname
  mn@eprint@\@tempb\endcsname \expandafter{\@tempc}}}

\bibitem[\protect\citeauthoryear{Abbott et~al.}{Abbott
  et~al.}{2017}]{TheLIGOScientific:2017qsa}
Abbott B.~P.,  et~al., 2017, \mn@doi [Phys. Rev. Lett.]
  {10.1103/PhysRevLett.119.161101}, 119, 161101

\bibitem[\protect\citeauthoryear{An et~al.}{An et~al.}{2020}]{AnZhengHua2022}
An Z.~H.,  et~al., 2020, \mn@doi [Radiation Detection Technology and Methods]
  {10.1007/s41605-021-00289-y}

\bibitem[\protect\citeauthoryear{Arnaud, Gordon  \& et al.}{Arnaud
  et~al.}{2022}]{Arnaud2022}
Arnaud K.,  Gordon C.,   et al. 2022, XSPEC Users’ Guide for version 12.12.1

\bibitem[\protect\citeauthoryear{Band et~al.}{Band et~al.}{1993}]{Band:1993}
Band D.,  et~al., 1993, \mn@doi [ApJ.] {10.1086/172995}, 413, 281

\bibitem[\protect\citeauthoryear{Blackburn, Briggs, Camp, Christensen,
  Connaughton, Jenke, Remillard  \& Veitch}{Blackburn
  et~al.}{2015}]{Blackburn:2014rqa}
Blackburn L.,  Briggs M.~S.,  Camp J.,  Christensen N.,  Connaughton V.,  Jenke
  P.,  Remillard R.~A.,   Veitch J.,  2015, \mn@doi [Astrophys. J. Suppl.]
  {10.1088/0067-0049/217/1/8}, 217, 8

\bibitem[\protect\citeauthoryear{Bochenek, Ravi, Belov, Hallinan, Kocz,
  Kulkarni  \& McKenna}{Bochenek et~al.}{2020}]{bochenek2020fast}
Bochenek C.~D.,  Ravi V.,  Belov K.~V.,  Hallinan G.,  Kocz J.,  Kulkarni
  S.~R.,   McKenna D.~L.,  2020, \mn@doi [Nature] {10.1038/s41586-020-2872-x},
  587, 59–62

\bibitem[\protect\citeauthoryear{{CHIME/FRB Collaboration} et~al.,}{{CHIME/FRB
  Collaboration} et~al.}{2020}]{2020Natur.587...54C}
{CHIME/FRB Collaboration} et~al., 2020, \mn@doi [\nat]
  {10.1038/s41586-020-2863-y}, \href
  {https://ui.adsabs.harvard.edu/abs/2020Natur.587...54C} {587, 54}

\bibitem[\protect\citeauthoryear{Cai et~al.}{Cai
  et~al.}{2021a}]{10.1093/mnras/stab2760}
Cai C.,  et~al., 2021a, \mn@doi [Monthly Notices of the Royal Astronomical
  Society] {10.1093/mnras/stab2760}

\bibitem[\protect\citeauthoryear{{Cai} et~al.,}{{Cai}
  et~al.}{2021b}]{2021GCN.30140....1C}
{Cai} C.,  et~al., 2021b, GRB Coordinates Network, \href
  {https://ui.adsabs.harvard.edu/abs/2021GCN.30140....1C} {30140, 1}

\bibitem[\protect\citeauthoryear{Cai et~al.,}{Cai
  et~al.}{2022}]{cai2022insighthxmt}
Cai C.,  et~al., 2022, Insight-HXMT dedicated 33-day observation of SGR
  J1935+2154 I. Burst Catalog (\mn@eprint {arXiv} {2203.16855})

\bibitem[\protect\citeauthoryear{{Collazzi} et~al.,}{{Collazzi}
  et~al.}{2015}]{2015ApJS..218...11C}
{Collazzi} A.~C.,  et~al., 2015, \mn@doi [\apjs] {10.1088/0067-0049/218/1/11},
  \href {https://ui.adsabs.harvard.edu/abs/2015ApJS..218...11C} {218, 11}

\bibitem[\protect\citeauthoryear{{Fishman} et~al.,}{{Fishman}
  et~al.}{1994}]{1994STIN...9611316F}
{Fishman} G.~J.,  et~al., 1994, {Discovery of intense gamma-ray flashes of
  atmospheric origin}, NASA STI/Recon Technical Report N

\bibitem[\protect\citeauthoryear{Fletcher}{Fletcher}{2021}]{2021GCN.30125....1F}
Fletcher C.,  2021, GRB Coordinates Network, \href
  {https://ui.adsabs.harvard.edu/abs/2021GCN.30125....1F} {30125, 1}

\bibitem[\protect\citeauthoryear{Goldstein, Burns, Hamburg, Connaughton, Veres,
  Briggs  \& Hui}{Goldstein et~al.}{2016}]{Goldstein:2016zfh}
Goldstein A.,  Burns E.,  Hamburg R.,  Connaughton V.,  Veres P.,  Briggs
  M.~S.,   Hui C.~M.,  2016, arXiv:1612.02395.

\bibitem[\protect\citeauthoryear{Goldstein et~al.}{Goldstein
  et~al.}{2017}]{Goldstein:2017mmi}
Goldstein A.,  et~al., 2017, \mn@doi [Astrophys. J.]
  {10.3847/2041-8213/aa8f41}, 848, L14

\bibitem[\protect\citeauthoryear{Goldstein et~al.}{Goldstein
  et~al.}{2019}]{Goldstein:2019tfz}
Goldstein A.,  et~al., 2019, arXiv:1903.12597

\bibitem[\protect\citeauthoryear{Guidorzi et~al.,}{Guidorzi
  et~al.}{2020a}]{Guidorzi_2020}
Guidorzi C.,  et~al., 2020a, \mn@doi [Astronomy \& Astrophysics]
  {10.1051/0004-6361/202037797}, 637, A69

\bibitem[\protect\citeauthoryear{Guidorzi et~al.,}{Guidorzi
  et~al.}{2020b}]{2020Guidorzi}
Guidorzi C.,  et~al., 2020b, \mn@doi [Astronomy \& Astrophysics]
  {10.1051/0004-6361/202039129}, 642, A160

\bibitem[\protect\citeauthoryear{{Hamburg} \& others"}{{Hamburg} \&
  others"}{2020}]{2020ApJ...893..100H}
{Hamburg} R.,  others" 2020, \mn@doi [\apj] {10.3847/1538-4357/ab7d3e}, \href
  {https://ui.adsabs.harvard.edu/abs/2020ApJ...893..100H} {893, 100}

\bibitem[\protect\citeauthoryear{{Hannam} \& {Thompson}}{{Hannam} \&
  {Thompson}}{1999}]{1999NIMPA.431..239H}
{Hannam} M.~D.,  {Thompson} W.~J.,  1999, \mn@doi [Nuclear Instruments and
  Methods in Physics Research A] {10.1016/S0168-9002(99)00269-7}, \href
  {https://ui.adsabs.harvard.edu/abs/1999NIMPA.431..239H} {431, 239}

\bibitem[\protect\citeauthoryear{{Hauschild} \& {Jentschel}}{{Hauschild} \&
  {Jentschel}}{2001}]{2001NIMPA.457..384H}
{Hauschild} T.,  {Jentschel} M.,  2001, \mn@doi [Nuclear Instruments and
  Methods in Physics Research A] {10.1016/S0168-9002(00)00756-7}, \href
  {https://ui.adsabs.harvard.edu/abs/2001NIMPA.457..384H} {457, 384}

\bibitem[\protect\citeauthoryear{{Hurley} et~al.,}{{Hurley}
  et~al.}{2005}]{2005Natur.434.1098H}
{Hurley} K.,  et~al., 2005, \mn@doi [\nat] {10.1038/nature03519}, \href
  {https://ui.adsabs.harvard.edu/abs/2005Natur.434.1098H} {434, 1098}

\bibitem[\protect\citeauthoryear{{Kaastra}}{{Kaastra}}{2017}]{2017A&A...605A..51K}
{Kaastra} J.~S.,  2017, \mn@doi [\aap] {10.1051/0004-6361/201629319}, \href
  {https://ui.adsabs.harvard.edu/abs/2017A&A...605A..51K} {605, A51}

\bibitem[\protect\citeauthoryear{Kocevski et~al.}{Kocevski
  et~al.}{2018}]{Kocevski:2018suj}
Kocevski D.,  et~al., 2018, \mn@doi [Astrophys. J.] {10.3847/1538-4357/aacb7b},
  862, 152

\bibitem[\protect\citeauthoryear{{Li} \& {Ma}}{{Li} \&
  {Ma}}{1983}]{1983ApJ...272..317L}
{Li} T.~P.,  {Ma} Y.~Q.,  1983, \mn@doi [\apj] {10.1086/161295}, \href
  {https://ui.adsabs.harvard.edu/abs/1983ApJ...272..317L} {272, 317}

\bibitem[\protect\citeauthoryear{{Li}, {Zhang}  \& {L{\"u}}}{{Li}
  et~al.}{2016}]{2016ApJS..227....7L}
{Li} Y.,  {Zhang} B.,   {L{\"u}} H.-J.,  2016, \mn@doi [\apjs]
  {10.3847/0067-0049/227/1/7}, \href
  {https://ui.adsabs.harvard.edu/abs/2016ApJS..227....7L} {227, 7}

\bibitem[\protect\citeauthoryear{Li et~al.}{Li et~al.}{2018}]{Li:2017iup}
Li T.,  et~al., 2018, \mn@doi [Sci. China Phys. Mech. Astron.]
  {10.1007/s11433-017-9107-5}, 61, 031011

\bibitem[\protect\citeauthoryear{{Li} et~al.,}{{Li}
  et~al.}{2021}]{2021NatAs...5..378L}
{Li} C.~K.,  et~al., 2021, Nature Astronomy, \href
  {https://ui.adsabs.harvard.edu/abs/2021NatAs...5..378L} {5, 378}

\bibitem[\protect\citeauthoryear{Li et~al.}{Li et~al.}{2022}]{LiXinQiao2022}
Li X.~Q.,  et~al., 2022, \mn@doi [Radiation Detection Technology and Methods]
  {10.1007/s41605-021-00288-z}

\bibitem[\protect\citeauthoryear{{Lin}, {Kouveliotou}  \& {van der
  Horst}}{{Lin} et~al.}{2011}]{2011AIPC.1358..313L}
{Lin} L.,  {Kouveliotou} C.,   {van der Horst} A.~J.,  2011, in {McEnery}
  J.~E.,  {Racusin} J.~L.,   {Gehrels} N.,  eds,  American Institute of Physics
  Conference Series Vol. 1358, Gamma Ray Bursts 2010. pp 313--316 (\mn@eprint
  {arXiv} {1107.2121}), \mn@doi{10.1063/1.3621796}

\bibitem[\protect\citeauthoryear{Lin, Gö{\u{g}}ü{\c{s}}, Roberts, Baring,
  Kouveliotou, Kaneko, van~der Horst  \& Younes}{Lin et~al.}{2020}]{Lin_2020b}
Lin L.,  Gö{\u{g}}ü{\c{s}} E.,  Roberts O.~J.,  Baring M.~G.,  Kouveliotou
  C.,  Kaneko Y.,  van~der Horst A.~J.,   Younes G.,  2020, \mn@doi [The
  Astrophysical Journal] {10.3847/2041-8213/abbefe}, 902, L43

\bibitem[\protect\citeauthoryear{{Liu} et~al.,}{{Liu}
  et~al.}{2020}]{2020SCPMA..6349503L}
{Liu} C.,  et~al., 2020, \mn@doi [Science China Physics, Mechanics, and
  Astronomy] {10.1007/s11433-019-1486-x}, \href
  {https://ui.adsabs.harvard.edu/abs/2020SCPMA..6349503L} {63, 249503}

\bibitem[\protect\citeauthoryear{{Lorimer}, {Bailes}, {McLaughlin}, {Narkevic}
  \& {Crawford}}{{Lorimer} et~al.}{2007}]{2007Sci...318..777L}
{Lorimer} D.~R.,  {Bailes} M.,  {McLaughlin} M.~A.,  {Narkevic} D.~J.,
  {Crawford} F.,  2007, \mn@doi [Science] {10.1126/science.1147532}, \href
  {https://ui.adsabs.harvard.edu/abs/2007Sci...318..777L} {318, 777}

\bibitem[\protect\citeauthoryear{Meegan et~al.,}{Meegan
  et~al.}{2009}]{Meegan_2009}
Meegan C.,  et~al., 2009, \mn@doi [The Astrophysical Journal]
  {10.1088/0004-637x/702/1/791}, 702, 791

\bibitem[\protect\citeauthoryear{Mereghetti et~al.,}{Mereghetti
  et~al.}{2020}]{Mereghetti_2020}
Mereghetti S.,  et~al., 2020, \mn@doi [The Astrophysical Journal]
  {10.3847/2041-8213/aba2cf}, 898, L29

\bibitem[\protect\citeauthoryear{{Moss}, {Lien}, {Guiriec}, {Cenko}  \&
  {Sakamoto}}{{Moss} et~al.}{2022}]{2022ApJ...927..157M}
{Moss} M.,  {Lien} A.,  {Guiriec} S.,  {Cenko} S.~B.,   {Sakamoto} T.,  2022,
  \mn@doi [\apj] {10.3847/1538-4357/ac4d94}, \href
  {https://ui.adsabs.harvard.edu/abs/2022ApJ...927..157M} {927, 157}

\bibitem[\protect\citeauthoryear{Neyman \& Pearson}{Neyman \&
  Pearson}{1928}]{10.1093/biomet/20A.3-4.263}
Neyman J.,  Pearson E.~S.,  1928, \mn@doi [Biometrika]
  {10.1093/biomet/20A.3-4.263}, 20A, 263

\bibitem[\protect\citeauthoryear{Qiao et~al.}{Qiao et~al.}{2022}]{QiaoRui2022}
Qiao R.,  et~al., 2022, \mn@doi [submitted to RAA] {10.1360/SSPMA-2019-0417}

\bibitem[\protect\citeauthoryear{{Ridnaia} et~al.,}{{Ridnaia}
  et~al.}{2021}]{2021NatAs...5..372R}
{Ridnaia} A.,  et~al., 2021, \mn@doi [Nature Astronomy]
  {10.1038/s41550-020-01265-0}, \href
  {https://ui.adsabs.harvard.edu/abs/2021NatAs...5..372R} {5, 372}

\bibitem[\protect\citeauthoryear{Savchenko et~al.}{Savchenko
  et~al.}{2017}]{Savchenko:2017ffs}
Savchenko V.,  et~al., 2017, \mn@doi [Astrophys. J.]
  {10.3847/2041-8213/aa8f94}, 848, L15

\bibitem[\protect\citeauthoryear{{Song} et~al.,}{{Song}
  et~al.}{2022}]{2022ApJS..259...46S}
{Song} X.-Y.,  et~al., 2022, \mn@doi [\apjs] {10.3847/1538-4365/ac4d22}, \href
  {https://ui.adsabs.harvard.edu/abs/2022ApJS..259...46S} {259, 46}

\bibitem[\protect\citeauthoryear{Tavani et~al.,}{Tavani
  et~al.}{2020}]{tavani2020xray}
Tavani M.,  et~al., 2020, An X-Ray Burst from a Magnetar Enlightening the
  Mechanism of Fast Radio Bursts (\mn@eprint {arXiv} {2005.12164})

\bibitem[\protect\citeauthoryear{{Xiao} et~al.,}{{Xiao}
  et~al.}{2022a}]{2022MNRAS.tmp..994X}
{Xiao} S.,  et~al., 2022a, \mn@doi [\mnras] {10.1093/mnras/stac999}, \href
  {https://ui.adsabs.harvard.edu/abs/2022MNRAS.tmp..994X} {}

\bibitem[\protect\citeauthoryear{Xiao et~al.,}{Xiao
  et~al.}{2022b}]{10.1093/mnras/stac085}
Xiao S.,  et~al., 2022b, \mn@doi [Monthly Notices of the Royal Astronomical
  Society] {10.1093/mnras/stac085}, 511, 964

\bibitem[\protect\citeauthoryear{Xiong et~al.,}{Xiong
  et~al.}{2012}]{articleXiong}
Xiong S.,  et~al., 2012, \mn@doi [Journal of Geophysical Research]
  {10.1029/2011JA017085}, 117, 1

\bibitem[\protect\citeauthoryear{Xiong et~al.}{Xiong
  et~al.}{2022}]{XiongShaoLin2022}
Xiong S.~L.,  et~al., 2022, \mn@doi [submitted to RAA]
  {10.1360/SSPMA-2019-0417}

\bibitem[\protect\citeauthoryear{{Yang} et~al.,}{{Yang}
  et~al.}{2020}]{2020ApJ...899..106Y}
{Yang} J.,  et~al., 2020, \mn@doi [\apj] {10.3847/1538-4357/aba745}, \href
  {https://ui.adsabs.harvard.edu/abs/2020ApJ...899..106Y} {899, 106}

\bibitem[\protect\citeauthoryear{Zhang, Li, Lu, Song, Xu, Liu  \& et al.}{Zhang
  et~al.}{2020}]{Zhang_2020}
Zhang S.-N.,  Li T.,  Lu F.,  Song L.,  Xu Y.,  Liu C.,   et al. 2020, \mn@doi
  [Science China Physics, Mechanics \& Astronomy] {10.1007/s11433-019-1432-6},
  63

\makeatother
\end{thebibliography}








\bsp	
\label{lastpage}
\end{document}